\documentclass[%
 reprint,
superscriptaddress,
 amsmath,amssymb,
 aps,
 pra
]{revtex4-2}

\usepackage{graphicx}
\usepackage{dcolumn}
\usepackage{bm}
\usepackage{soul}
\usepackage[colorlinks]{hyperref}

\usepackage[english]{babel}
\usepackage{color}
\usepackage{mathtools}
\usepackage{cleveref}
\usepackage{verbatim}
\usepackage{breqn}
\usepackage[dvipsnames]{xcolor}
\makeatletter
\newcommand{\customlabel}[2]{%
   \protected@write \@auxout {}{\string \newlabel {#1}{{#2}{\thepage}{#2}{#1}{}} }%
   \hypertarget{#1}{}
}
\makeatother

\begin{document}

\title{Self-pulsing and chaos in the asymmetrically-driven dissipative photonic Bose-Hubbard dimer: A bifurcation analysis}

\author{Jes\'us Yelo-Sarri\'on}
\author{Francois Leo}
\author{Simon-Pierre Gorza}
\affiliation{OPERA-Photonique$,$ Université libre de Bruxelles$,$ 50 Avenue F. D. Roosevelt$,$ CP 194/5 B-1050 Bruxelles$,$ Belgium}

\author{Pedro Parra-Rivas}

\affiliation{OPERA-Photonique$,$ Université libre de Bruxelles$,$ 50 Avenue F. D. Roosevelt$,$ CP 194/5 B-1050 Bruxelles$,$ Belgium}

\affiliation{Dipartimento di Ingegneria dell’Informazione$,$ Elettronica e Telecomunicazioni$,$
Sapienza Universit\'a di Roma$,$ via Eudossiana 18$,$ 00184 Rome$,$ Italy\\}

\date{\today}

\begin{abstract}
We perform a systematic study of the temporal dynamics emerging in the asymmetrically driven dissipative Bose-Hubbard dimer model. This model successfully describes the nonlinear dynamics of photonic diatomic molecules in linearly coupled Kerr resonators coherently excited by a single laser beam.
Such temporal dynamics include self-pulsing oscillations, period doubled oscillatory states, chaotic dynamics, and spikes. The different states and dynamical regimes have been thoroughly characterized using bifurcation analysis. This analysis has allowed us to identify the main instabilities, i.e. bifurcations, responsible for the appearance of the previously stated dynamics.  

\end{abstract}

\maketitle
    
\section{Introduction}

The emergence of self-sustained oscillations is commonly encountered in a variety of different fields ranging from chemistry and biology to physics and engineering \cite{Nicolis1995,Jenkins2013}. Some examples include the Belousov-Zhabotinsky reaction \cite{zhabotinsky_history_1991,epstein_introduction_1998}, pulsation in Cepheid variable stars \cite{Jenkins2013}, or the oscillatory biochemical dynamics responsible for the cell division cycle \cite{pomerening_building_2003}. In all these systems, the self-pulsing behavior emerges when varying a given parameter beyond a critical value, where time translation symmetry is broken. This oscillatory instability is known as a Hopf bifurcation \cite{guckenheimer_nonlinear_1983, glendinning_stability_1994} and is key to understanding these dynamical phenomena. The self-sustained oscillations can evolve into much more complex dynamics. One example is chaos \cite{ott_chaos_2002}, which is present in different natural systems including weather and climate, fluid turbulent flow \cite{lorenz_deterministic_1963}, and chemical reactions \cite{petrov_controlling_1993}, to cite only a few.


Such rich temporal dynamics may also appear in optical systems. For example, self-pulsing has been found in second-harmonic generation \cite{Drummond1980}, in lasers with continuous injected signals \cite{Lugiato1983} or between coupled longitudinal modes \cite{Paoli1969} and, more recently, 
in coupled photonic cavities \cite{Maes2009,Petracek2014,Grigoriev2011,Sato2012} or between counter-propagating beams in single Kerr cavities \cite{Woodley2020}. Furthermore, period-doubling and chaos have also been demonstrated (see e.g. Refs \cite{Bessin2019, Virte2013}).

Lately, different works have focused on the dynamics of two coupled Kerr resonators: for the generation of optical frequency combs in integrated ring resonators in the anomalous \cite{Tikan2020} or the normal \cite{Xue2015, Fujii2018} dispersion regime, as well as for parametric oscillation generation in coupled polariton \cite{Zambon2020} or fiber \cite{ yelo-sarrion_self-pulsing_2021} cavities.  
This system is commonly known as {\it driven dissipative photonic Bose-Hubbard dimer} (PBHD) \cite{Abbarchi2013}, in analogy with the open quantum boson system \cite{Bruder2005}.
Advances on those different platforms have a broad range of applications ranging from spectroscopy \cite{Kowligy2019} or metrology \cite{Newman2019} to all-optical information storage \cite{Leo2010}. Moreover, the addition of a second cavity has until now been proven useful to describe physical systems at the molecular level as shown by Zhang et al. \cite{Zhang2019}. Those levels of control over tunable photonic systems pave the way to major advances in new unconventional information processing techniques such as all-optical \cite{Feldmann2019} or quantum computing \cite{Kues2017}.

In a previous work, we found that self-pulsing oscillations appear in PBHDs when the driving is asymmetrical, i.e., when only one cavity is driven \cite{yelo-sarrion_self-pulsing_2021}.  There, we focused on the self-pulsing dynamics, and we showed an excellent agreement between the experimental observations in coupled fiber cavities (in the normal dispersion regime) and the theoretical modeling. However, akin to Refs \cite{Giraldo2020, Petracek2014}, much richer dynamical regimes ranging from period doubled oscillatory states, temporal chaos, and the presence of spikes (i.e. homoclinic orbits) are expected.
Thus, the main aim of this paper is to perform a systematic study of the dynamics, stability, and bifurcation structure of asymmetrically-driven dissipative PBHDs, applying well-known techniques of dynamical systems and bifurcation theory.

\begin{figure}
\includegraphics[width=\columnwidth]{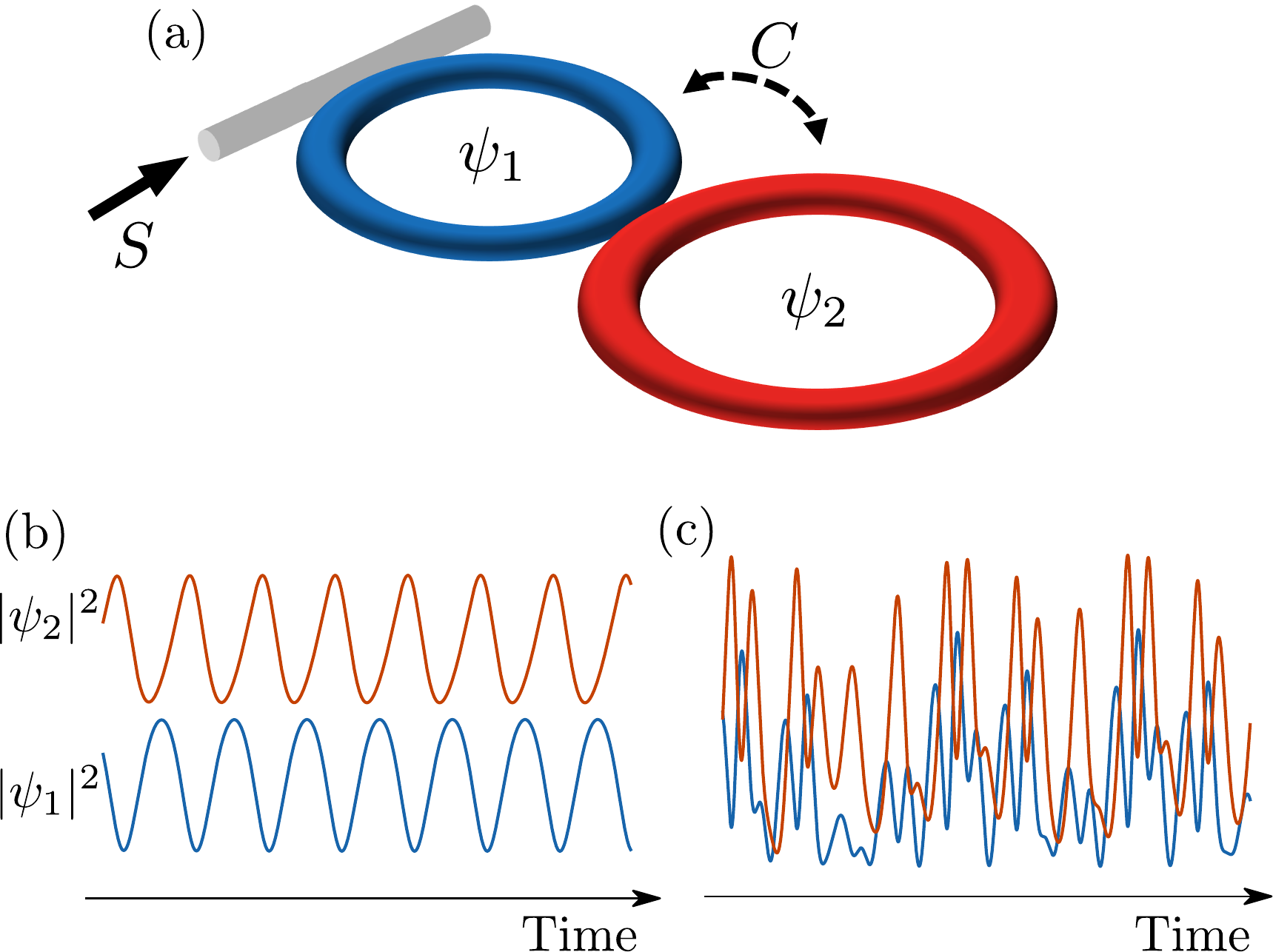}
\caption{(a) Schematic view of an asymmetrically-driven dissipative PBHD. $C$ represents the coupling between cavities and $S$ the driving field. (b) Temporal traces of the cavity intensities $|\psi_{1,2}|^2$ in a self-pulsing regime. (c) Same as in (b) but in a chaotic dynamics regime.}  
\label{fig0}
\end{figure}



This article is organized as follows. In Section~\ref{sec:1} we introduce the driven dissipative PBHD model, and we describe some of its characteristics, the 4D dynamical system associated with it, and the methodological approach that we will follow. Section~\ref{sec:2} focuses on the steady states of the system, analyzing their linear stability and steady bifurcations for two coupling regimes. After that, in Section~\ref{sec:3} we perform a detailed bifurcation analysis of the previous coupling regimes, and we characterize the different dynamical scenarios. The origin of such dynamics is related to the presence of Hopf and homoclinic bifurcations. Later (see Section~\ref{sec:4}), we study two homoclinic orbits of different types and present their main features. Section~\ref{sec:5} is devoted to elucidating the origin of the temporal chaos found in the system. Finally, we discuss our results and draw our conclusions in Section~\ref{sec:6}.

\section{The asymmetrically-driven dissipative photonic Bose-Hubbard dimer model}\label{sec:1}
Let us assume two identical cavities, mutually coupled via a middle coupler and coherently driven through a single input coupler [see Fig.\,\ref{fig0}(a)]. Regarding experiments, this scheme may be easier than with two pumps for which the relative phase between them must be precisely controlled. We here consider one-dimensional coupled cavities such as micropillar or photonic crystal cavities, but also coupled ring resonators (see discussion in Ref \cite{yelo-sarrion_self-pulsing_2021} regarding the dispersion). The dynamics of such PBHD can be described by two coupled one-dimensional normalized Lugiato-Lefever equations \cite{Lugiato1987}
\begin{equation}
\begin{array}{lll}
    \displaystyle \frac{d \psi_1}{dt} &=&\displaystyle [-1 +i(|\psi_1|^2 - \Delta)]\psi_1 + iC\psi_2+S\\
    \\
    \displaystyle \frac{d \psi_2}{dt} &=& \displaystyle[-1 +i(|\psi_2|^2 - \Delta)]\psi_2  + iC\psi_1
\label{eq-coupled-lle-diff}
\end{array}
\end{equation}
where the time $t=t'\kappa/T_R$ with $t'$ the laboratory time, $T_R$ the round-trip time and $\kappa$ the cavity loss coefficient (excluding the middle coupler). The detuning from the closest (single cavity) resonances is $\Delta=\delta/\kappa=(m2\pi-\varphi)/\kappa$, where $\varphi$ is the round-trip linear phase shift and $m$ an integer number. $C=\sqrt{\theta_{12}}/\kappa$, where $\theta_{12}$ is the transmission coefficient of the coupler between the cavities. Finally $\psi_j=A_j\sqrt{\gamma L / \kappa}$, ($j=1,~2$), $S=i\sqrt{P_p\gamma L \theta_p/ \kappa^3}$, where $A_j$ are the field amplitudes normalized such that the intracavity powers (expressed in watts) are given by $|A_j|^2=P_j$. $P_p$ is the driving power, $\theta_p$ is the transmission coefficient of the input coupler, $\gamma$ is the nonlinear parameter of the waveguide and $L$ is the length of the resonators.
%
The normalization scheme follows the one proposed in Ref.~\cite{Leo2010}. 


In order to study this system we are going to apply two different procedures: the direct numerical integration of Eqs.~(\ref{eq-coupled-lle-diff}), using a split-step algorithm, and the numerical path-continuation \cite{DoedelI,DoedelII} of the different states of the system using the free software continuation package AUTO-07p \cite{Doedel2009}. This last approach allows us to characterize the bifurcation structure and stability of the different static and periodic dynamical states as a function of the parameters of the system. For the  periodic states, Floquet analysis is applied  \citep{wiggins_introduction_2003}.

To perform numerical parameter continuation, it is convenient to recast the complex dynamical system (\ref{eq-coupled-lle-diff}) into the 4D system 
\begin{equation}
\begin{array}{lll}
    \displaystyle \frac{d u_1}{d t} &=& -u_1  - (u_1^2 + v_1^2 - \Delta) v_1- C v_2+S\\
    \\
    \displaystyle \frac{d v_1}{d t} &=& -v_1  + (u_1^2 + v_1^2 - \Delta) u_1 + C u_2\\
    \\
    \displaystyle \frac{d u_2}{d t} &=& -u_2  - (u_2^2 + v_2^2 - \Delta) v_2 - C v_1\\
    \\
    \displaystyle \frac{d v_2}{d t} &=& -v_2  + (u_2^2 + v_2^2 - \Delta) u_2 + Cu_1,\\
\label{system-lle-hss}
\end{array}
\end{equation}
where $\psi_j = u_j + iv_j$, for $j=1,2$. As control parameters we consider the pump intensity $S$, the coupling constant $C$, and the phase detuning $\Delta$. 

This system supports self-pulsing dynamics, i.e., periodic oscillations in time, as schematically shown in Fig.~\ref{fig0}(b).
Through the modification of suitable parameters, these oscillations can suffer different transitions leading to more complex dynamics. One example of such complexity, corresponding to chaos, is depicted in Fig.~\ref{fig0}(c). In what follows we will unveil the features of these states and the transition that they may encounter.

\section{Steady-states, linear stability, and their phase diagram}\label{sec:2}

\begin{figure*}
\includegraphics{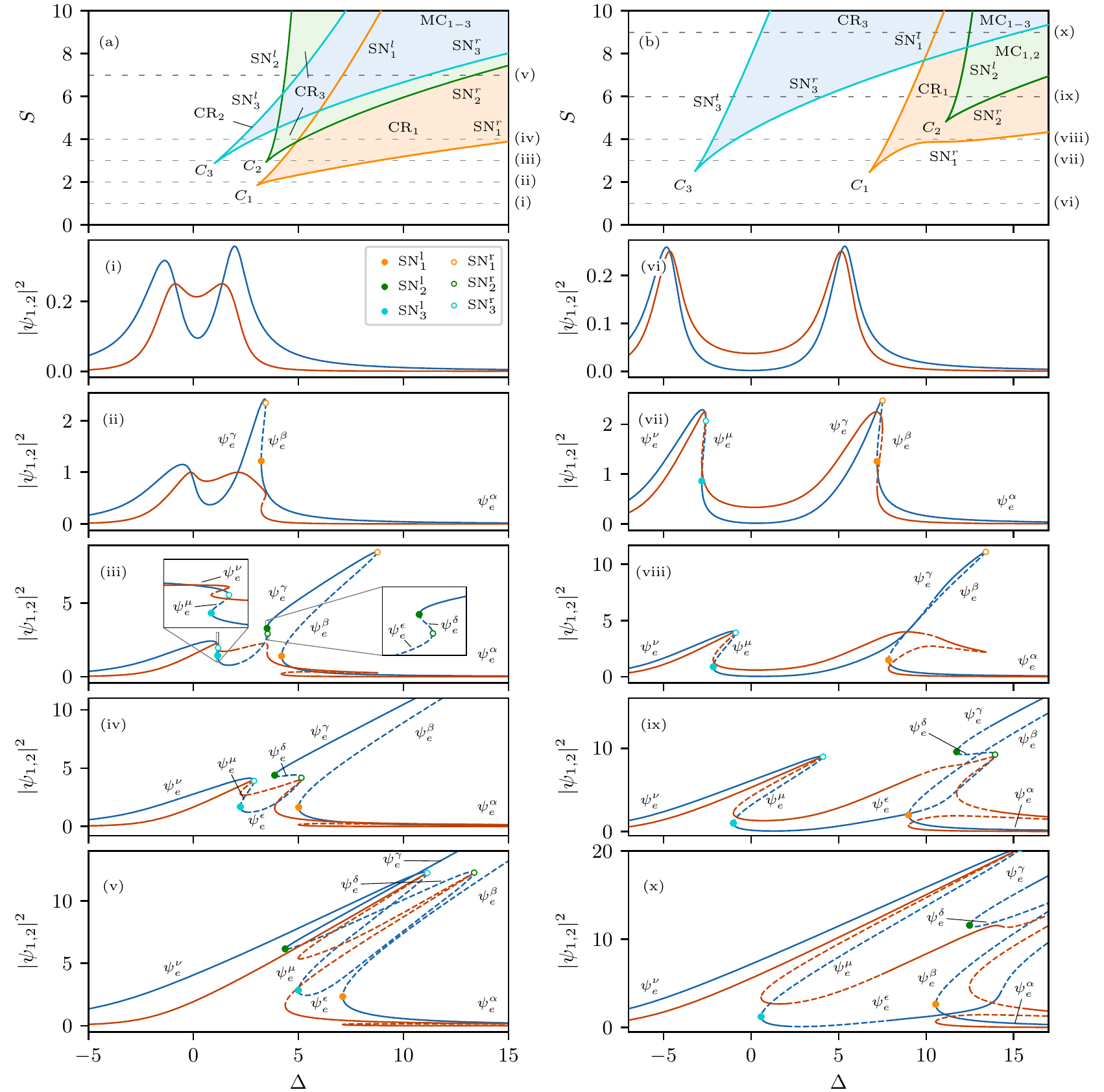}
\caption{Steady-state phase diagrams in the weakly and strongly couple regimes. (a) shows the $(\Delta,S)$-phase diagram for $C=1.5$. The bifurcation diagrams shown in panels (i)-(iv) correspond to slices of constant $S$ [see dashed horizontal lines in (a)]. They represent the intensities $|\psi_{1}|^2$ (blue) and $|\psi_{2}|^2$ (red) as a function of $\Delta$. From top to bottom the values are $S = 1,2,3,4$ and 7.
(b) shows the $(\Delta,S)$-phase diagram for $C=5$. Panels (v)-(viii) are slices of panel (b) for the constant values 
$S = 1,3,4,6$ and 9 (from top to bottom).
Solid and dashed lines represent stable and unstable equilibria ($\psi_e^{i}$) respectively. In (a) and (b) different coexistence regions (CR$_j$), and multi-coexistence regions (MC$_j$) are depicted as well as the static bifurcations of the equilibria: saddle-node SN$_i^{l,r}$ and cusp $C_i$.
}
\label{fig1}
\end{figure*}

The steady-states of the system correspond to the fixed points or equilibria $\psi_e=(\psi_1^e,\psi_2^e)=(u_1^e,v_1^e,u_2^e,v_2^e)$ of equations (\ref{eq-coupled-lle-diff}), such that $d\psi/dt=0$. They are solution of the algebraic system
\begin{equation}
\begin{array}{lll}
    \displaystyle [-1 +i(|\psi_1|^2 - \Delta) ]\psi_1+ iC\psi_2+S&=&0\\
    \\
    \displaystyle [-1 +i(|\psi_2|^2 - \Delta) ]\psi_2+ iC\psi_1&=&0\\
\label{eq-coupled-lle-hss}
\end{array}
\end{equation}
which leads to
\begin{equation}
\begin{array}{lll}
    I_1 + (I_1-\Delta)^2I_1-C^2I_2-S^2+2CSv_2&=&0\\
    \\
    I_2 + (I_2-\Delta)^2I_2-C^2I_1&=&0\\
\label{hss_puissances}
\end{array}
\end{equation}
 The linear stability of these points is obtained from the local dynamics of (\ref{system-lle-hss}) around $\psi_e$, which is solely determined by the eigenvalues $\lambda$ of the Jacobian $\mathcal{J}$ of the system at that point.
%
The complicated form of Eq.~(\ref{hss_puissances}) prevents the possibility of extracting an analytical expression of the eigenvalues of the system. However, the equilibria and their stability can be easily computed numerically through the path-continuation algorithm \cite{DoedelI,DoedelII}. 

When two identical cavities are coupled, each resonance is split in two peaks. These peaks correspond to the excitation of the hybridized modes, often called the antibonding-like ($\Delta<0$) and the bonding-like ($\Delta>0$) modes of the photonic dimer. The detuning separation between them is equal to $2C$ and can thus be freely adjusted through the cavity coupling strength.  
In this study we consider two main dynamical regimes corresponding to the {\it weakly-coupled} scenario ($C=1.5$) for which the splitting is only slightly larger than the resonance width, and the {\it strongly-coupled} scenario ($C=5$), showing a much larger peak separation.
Hereafter we refer to them as WC and SC regimes respectively. The coupling constant that we consider in the WC regime is similar to the experimental value considered in our previous work \cite{yelo-sarrion_self-pulsing_2021}. 
In the SC, while the encountered dynamics are similar, the bifurcation diagram is much more complex.
In what follows, we study these two regimes separately. Their steady-state bifurcation structure is summarized in the diagrams plotted in Fig.~\ref{fig1}.

\subsection{Steady states in the weakly coupled regime}\label{sec:2.1}
Figure~\ref{fig1}(a) shows the steady-state phase diagram in the $(\Delta,S)$-parameter space for $C=1.5$. Slices of such diagram, for increasing values of $S$, are depicted in Figures~\ref{fig1}(i)-(iv) where the intensities $|\psi_{1,2}|^2$ are plotted against $\Delta$. The linear stability of these states is marked using solid (dashed) lines for stable (unstable) equilibria, and only steady-state bifurcations are labeled.

The two resonances are quite overlapped in the WC regime as can be seen for $S=1$ [see Fig.~\ref{fig1}(i)] for which the system is still close to the linear regime. At this driving level, the resonances are only slightly asymmetric, a single equilibrium exists for each given value of $\Delta$ and it is linearly stable.

Increasing $S$ [see Fig.~\ref{fig1}(ii) for $S=2$], the asymmetry between the resonances increases, yielding a larger right resonance in detriment of the left one. The right resonance undergoes a cusp or hysteresis bifurcation $C_1$, where a pair of folds, or turning points, are created, leading to the tilted shape. \cite{wiggins_introduction_2003}  These folds correspond to saddle-node bifurcations that we label SN$_{1}^{l,r}$. At these bifurcations, a stable node equilibrium and an unstable saddle collide and disappear. This transition leads to the coexistence of three different equilibria $\psi_e^{\alpha,\beta,\gamma}$, where $\psi_e^\alpha$ and $\psi_e^\gamma$ are nodes (i.e., stable), whereas $\psi_e^\beta$ is a saddle \cite{wiggins_introduction_2003}. The separation in $\Delta$ between SN$_1^l$ and SN$_1^r$ defines the {\it coexistence region}  CR$_1$ (see light orange area).

Further increasing $S$, both resonances tilt to the right due to the effect of the nonlinearity and two new cusp bifurcations occur [see Figs.~\ref{fig1}(a) and~\ref{fig1}(iii) for $S=3$]: $C_2$ on the right resonance, and $C_3$ on the left one. In $C_2$, SN$_{2}^{l,r}$ are created. These pair of bifurcations leads to the equilibria $\psi_e^\delta$ and $\psi_e^\epsilon$ (see close-up view) which are both unstable and coexist in CR$_2$. $C_3$ creates SN$_{3}^{l,r}$, and the two new equilibria $\psi_e^\mu$ and $\psi_e^\nu$ associated with the left resonance appear. Between SN$_{3}^{l,r}$ the bistability region CR$_3$ is created.

In Fig.~\ref{fig1}(iv) we plot the bifurcation diagram for $S=4$. For this value, the tilting of the resonance is much more prominent, and the bistability interval between $\psi_e^\alpha$ and $\psi_e^\gamma$ has increased considerably. For increasing values of $S$, the separation between the pairs SN$_{1}^{l,r}$, SN$_{2}^{l,r}$, SN$_{3}^{l,r}$ increases [see Fig.~\ref{fig1}(a)], and a {\it multi-coexistence region} appears between SN$_1^l$ and SN$_3^r$ that we label MC$_{1-3}$. An example of this situation is depicted in Fig.~\ref{fig1}(v) for $S=7$. Here we can see how both resonances are overlapped and lead to tristability and an effective single resonance.

\subsection{Steady states in the strongly coupled regime}\label{sec:2.2}
Let us now focus on a scenario with a higher coupling and fix $C=5$.
Figure~\ref{fig1}(b) shows the phase diagram for that coupling, and Figs.~\ref{fig1}(v)-(x) the bifurcation diagrams for some relevant values of $S$. In the linear regime the two resonances are now well separated. At $S=1$, they are still almost symmetric [see Fig.~\ref{fig1}(iv)]. 

Increasing $S$, the system encounters $C_1$ and $C_3$ almost simultaneously, where the pairs SN$_1^{l,r}$ and SN$_3^{l,r}$, together with the equilibria $\psi_e^{\alpha,\dots,\epsilon}$ are created as seen in Figs.~\ref{fig1}(b). 
These bifurcations define the coexistence regions CR$_1$ and CR$_2$.
An example of this configuration is plotted in Fig.~\ref{fig1}(vii) for $S=3$, where different equilibria are depicted. For this value of $S$, the resonance already tilts slightly to the right.

The configuration suffers an interesting modification for $S=4$ as shown in Fig.~\ref{fig1}(viii). For this value, the right resonance has just merged with an isola in a necking bifurcation (not shown here) \cite{Prat2002}. In this bifurcation, SN$_1^r$ and the left fold of the isola SN'$_1^l$ collide and the equilibria branches reconnect, enlarging in this way the arm of the right resonance which now extends until SN'$_1^r$ [relabelled  as SN$_1^r$ in Figs.~\ref{fig1}(a) and \ref{fig1}(viii) for clarity]. This type of merging is illustrated in detail in \cite{yelo-sarrion_self-pulsing_2021}.

Moving up in Fig.~\ref{fig1}(b), $C_2$ occurs and SN$_2^{l,r}$ are created. After this point, the configuration is like the one depicted in Fig.~\ref{fig1}(ix) for $S=6$. For this value of $S$, SN$_1^l$ and SN$_1^r$ are far apart, and the right resonance extends to larger values of $\Delta$.

For $S=9$ [see Fig.~\ref{fig1}(x)], the left resonance has grown significantly as SN$_3^r$ overpasses SN$_1^l$ and SN$_2^l$. At this stage, a tristable regime appears. Tristability persists for even larger values of $S$, where an effective single resonance can emerge similarly to the case shown in Fig.~\ref{fig1}(v). Here we can identify two main regions of multi coexistence that are labeled MC$_{1,2}$ and MC$_{1-3}$.

\section{Dynamical regimes and bifurcation structure}\label{sec:3}
In the previous section, we have focused on the steady-state equilibria and their steady bifurcations.  As previously stated, self-pulsing oscillations and chaos may emerge in the system [see Figs.~\ref{fig0}(b) and (c)]. In this section, we expand the previous analysis by studying the dynamical behavior of the system and present a systematic study of the bifurcation structure associated with such states in the WC and SC regimes. 
\begin{figure*}
\includegraphics{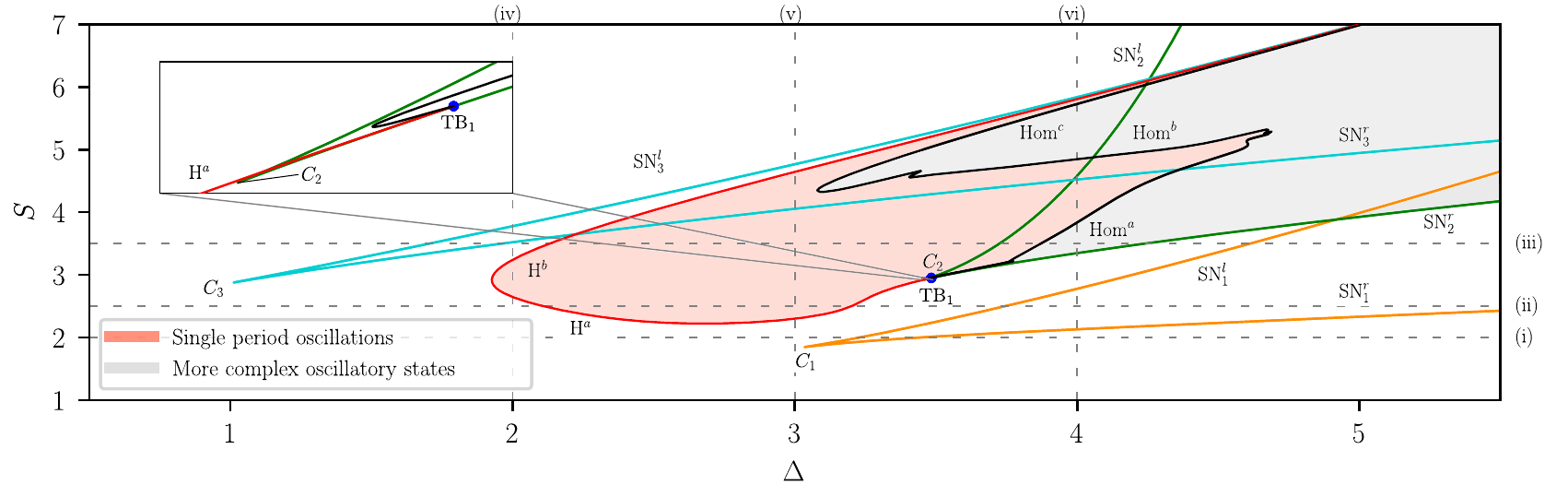}
\includegraphics{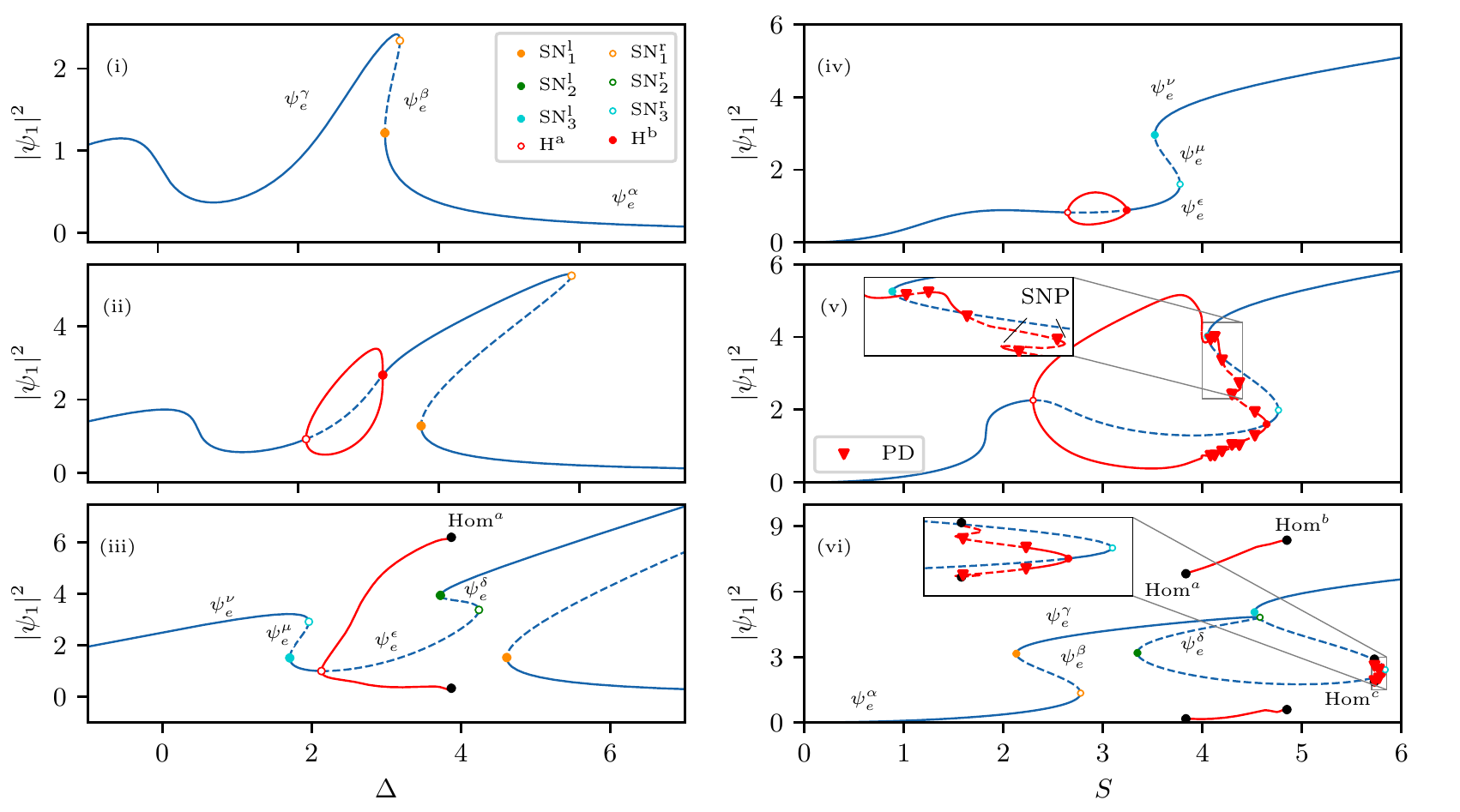}
\caption{(a) Phase diagram in the $(\Delta,S)$-parameter space for $C=1.5$ showing the main dynamical regions and bifurcations of the system: saddle-node of the steady state SN$_i^{l,r}$, cusp $C_i$, Hopf H$^{a,b}$, saddle-node of periodic oscillatory states SNP, homoclinic Hom$^{a,b}$, and Takens-Bogdanov TB$_j$.
The red shadowed region corresponds to self-pulsing dynamics. The vertical and horizontal dashed lines correspond to the bifurcation diagrams shown in panels (i)-(iii) for constant $S$ ($=2$, 2.5 and 3.5), and panels (iv)-(vi) for constant $\Delta$ ($= 2$, 3 and 4). Stable and unstable equilibria ($\psi_e^{i}$) are plotted with solid and dashed lines respectively. The red lines represent the maxima and minima of the periodic oscillations. 
}
\label{fig2}
\end{figure*}

\begin{figure*}
\includegraphics{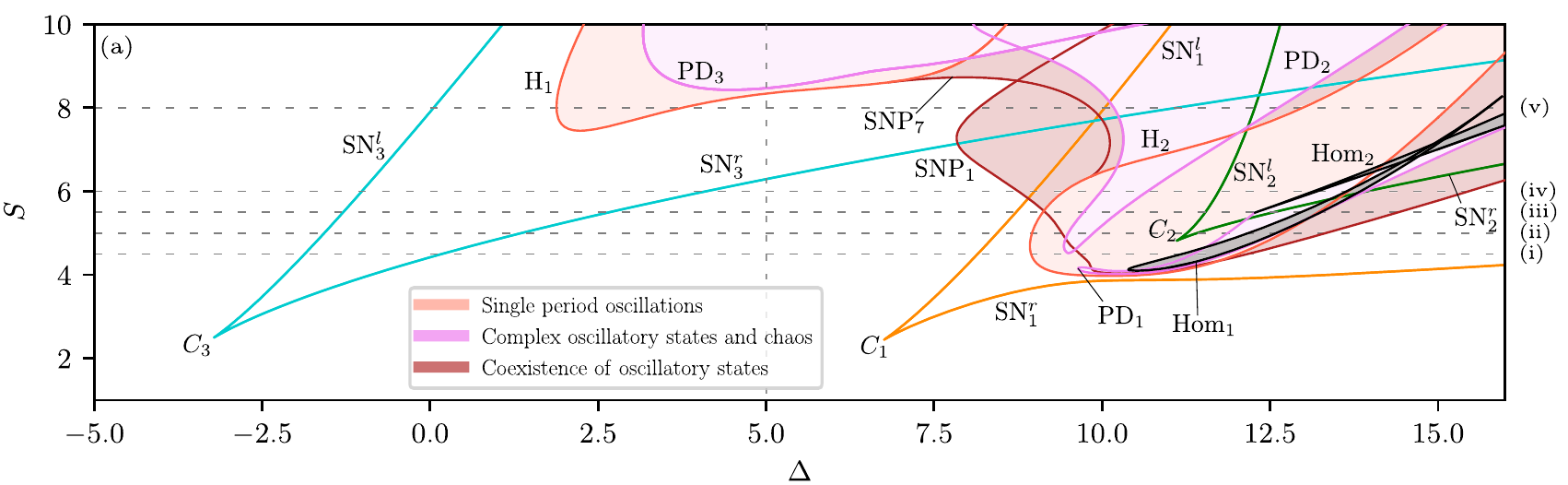}
\includegraphics{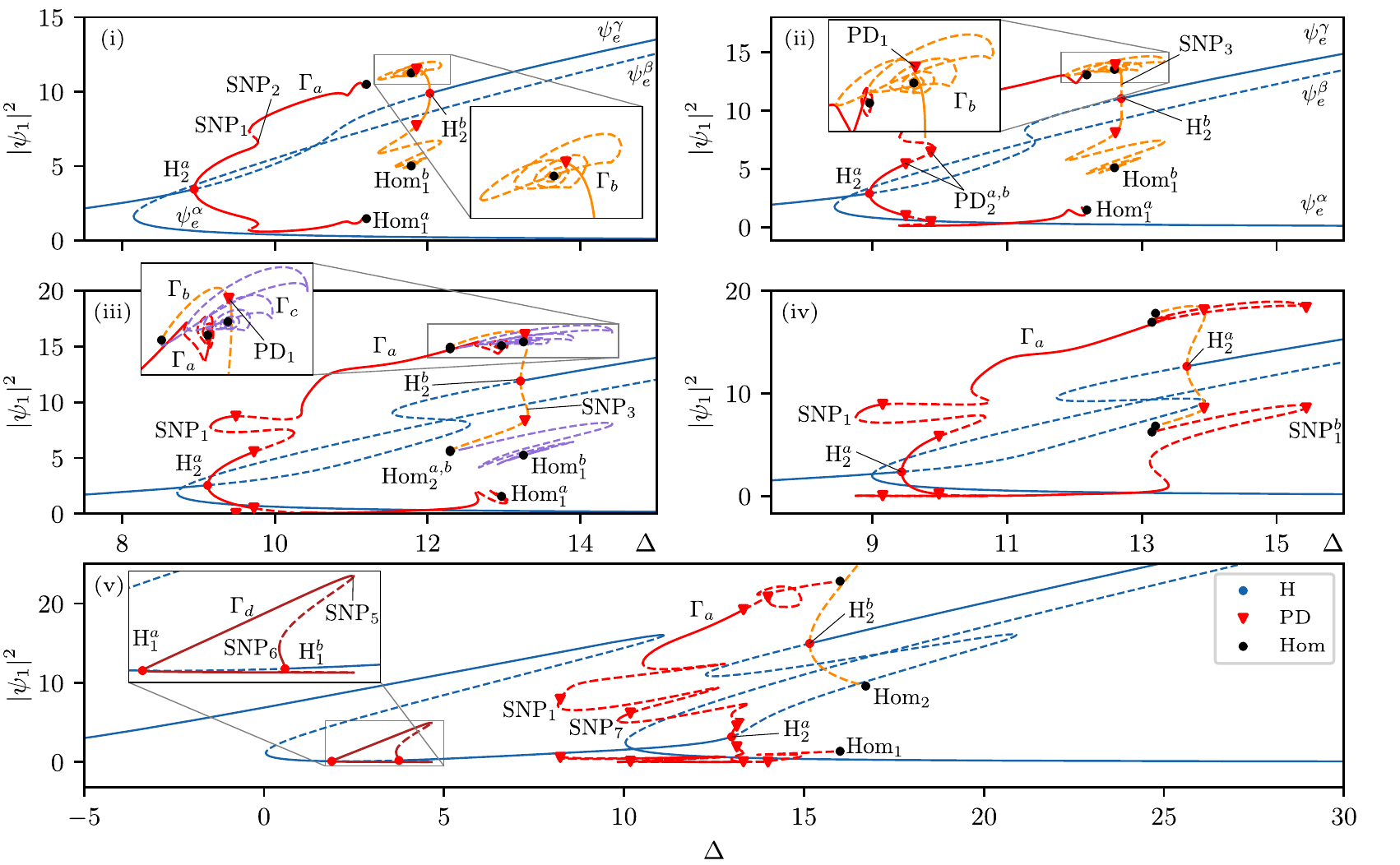}
\caption{Bifurcation structure for $C=5$. (a) shows the phase diagram in the $(\Delta, S)$-parameter space which illustrates the different dynamical regions and main bifurcations of the system: saddle-node of the steady-state SN$_i^{l,r}$, Hopf H$_{1,2}^{a,b}$, period-doubling PD$_i^{a,b}$, saddle-node of periodic oscillatory SNP$_i$, and homoclinic Hom$_{1,2}^{a,b}$. The horizontal dashed lines correspond to the bifurcation diagrams shown below, where $|\psi_1|^2$ is plotted as a function of $\Delta$ for $S=$ 4.5 (i), 5 (ii), 5.5 (iii), 6 (iV) and 8 (v). 
The vertical dashed line corresponds to the bifurcation diagram shown in Fig.~\ref{fig8}(a). In diagrams (i)-(v) solid (dashed) lines represent stable (unstable) equilibria ($\psi_e^{i}$) and limit cycles ($\Gamma_{i}$).  
}
\label{fig3}
\end{figure*}

\subsection{Dynamics in the weakly coupled regime}
The phase diagram shown at the top of Fig~\ref{fig2} summarizes the main dynamical regimes and bifurcation lines of the system for $C=1.5$. To understand such a diagram we slice it as shown by the horizontal and vertical dashed lines.
Each of these lines corresponds to one of the bifurcation diagrams shown below. The slices with constant $S$ are plotted in Figs.~\ref{fig2}(i)-(iii), whereas those with constant $\Delta$ are depicted in Fig~\ref{fig2}(iv)-(vi).

Figure~\ref{fig2}(i) shows a close-up view of Fig.~\ref{fig1}(ii) around the bistability region between $\psi_e^\alpha$ and $\psi_e^\gamma$ for $S=2$.  Upon increase of $S$, the Hopf bifurcation (H) line plotted in Fig.~\ref{fig2}(a) is crossed, and the system enters a self-pulsing regime (see red shadowed region) characterized by  single period oscillations like those shown in Fig.~\ref{fig0}(b). An example of this configuration is plotted in Fig.~\ref{fig2}(ii) for $S=2.5$, where H is crossed at two points H$^a$ and H$^b$. The periodic oscillations emerge supercritically from H$^a$ and with small amplitude from the left. The maximum and minimum of the oscillation are represented using red solid lines. Increasing $\Delta$, the oscillation amplitude grows until suddenly it dies out at H$^b$ on the right. This dramatic change of the oscillation amplitude in phase space is called {\it canard explosion} \cite{bold_forced_2003}, and is related to type II excitability \cite{izhikevich_neural_2012,parra-rivas_competition_2016}. 

For $S=3.5$, the situation is shown in Fig.~\ref{fig2}(iii). For this value, the equilibria  $\psi_e^\delta$, $\psi_e^\epsilon$, and $\psi_e^\mu$, $\psi_e^{\nu}$,
coexist. Furthermore, periodic oscillations persist. On the left, they still emerge from H$^a$. On the right, however, they die on a {\it homoclinic} (Hom) bifurcation \citep{glendinning_stability_1994,homburg_homoclinic_2010}, while only steady states persist for larger values of $\Delta$.

The homoclinic bifurcations are global bifurcations related to the collision of a cycle (i.e. a periodic orbit $\Gamma$) with an equilibrium, and are characterized by the divergence of the cycle's period \cite{homburg_homoclinic_2010}. At the bifurcation point, the limit cycle $\Gamma$ becomes a homoclinic orbit $\gamma$, i.e., a trajectory in the phase space which is bi-asymptotic to the equilibrium. This bifurcation is associated with type-I excitability \cite{izhikevich_neural_2012}.

In the phase diagram of Fig.~\ref{fig2}(a), the Hom bifurcation corresponds to the black solid line. To track numerically this line in the $(\Delta, S)$-parameter space we have used the homoclinic continuation HOMCONT extension of  AUTO-07p \cite{champneys_numerical_1996}.  We will focus our attention on this type of bifurcations in Sec.~\ref{sec:4}. 

We can also analyze the $(\Delta,S)$-phase diagram considering slices of constant $\Delta$. The resulting bifurcation diagrams are shown in Figs.~\ref{fig2}(iv)-(vi), where $|\psi_1|^2$ is plotted as a function of $S$. In Fig.~\ref{fig2}(iv) [$\Delta=2$] bistability exist between SN$_3^l$ and SN$_3^r$, and the limit cycle $\Gamma$ emerges and dies at H$^a$ and H$^b$ respectively. For this slice, bistability exists between $\psi_e^\epsilon$ and $\psi_e^{\nu}$.

Increasing $\Delta$ [see Fig.~\ref{fig2}(v) for $\Delta=3$], the bistability interval is now bound by H$^b$ and SN$_3^r$. $\Gamma$ increases drastically its amplitude and undergoes several secondary bifurcations such as saddle-node bifurcation of periodic orbits (SNP), also known as fold of cycles, and period-doubling bifurcations (PD) \citep{glendinning_local_1984,wiggins_introduction_2003}. For clarity, we do not plot these bifurcation lines in the phase diagram of Fig.~\ref{fig1}(a).
The presence of a PD may suggest the existence of chaotic dynamics emerging from a period-doubling cascade \cite{ott_chaos_2002}. We will analyze the chaotic dynamics of this system in Sec.~\ref{sec:5}.

In Fig.~\ref{fig2}(vi) [$\Delta=4$] we plot the bifurcation diagram after the occurrence of $C_2$ and $C_1$. Here, the bistability between H$^b$ and SN$_3^r$ reduces drastically (see close-up view), and the pair of bifurcations SN$_{2}^{l,r}$ and  SN$_{1}^{l,r}$ appear. Between the last two bifurcations, a new bistability range emerges, where $\psi_e^\alpha$ and $\psi_e^\gamma$ coexist. This slice cuts Hom in three different points that we label Hom$^{a,b,c}$, respectively. Periodic oscillations exist between Hom$^a$ and Hom$^b$, and between Hom$^c$ and H$^b$. 

The region in-between the H and Hom lines in the $(\Delta, S)$-parameter space [see the red shadowed area in Fig.~\ref{fig2}(a)] is the dynamical region of the system where self-pulsing and other dynamical states (e.g., chaos) may emerge. 

The H and Hom bifurcations arise from a pair of codimension-two Takens-Bodganov (TB) bifurcations \cite{wiggins_introduction_2003,guckenheimer_nonlinear_1983} which occur at SN$_2^r$ and SN$_3^l$. For the range of parameters considered here, we only observe TB$_1$ [see a close-up view in Fig.~\ref{fig2}(a)]. At this bifurcation, the linearized dynamics of the system has two zero eigenvalues $\lambda_{1,2}=0$ (with algebraic multiplicity 2), for this reason, it is also known as a double zero bifurcation \citep{guckenheimer_nonlinear_1983}. The periodic oscillations arise from this point with an infinite period, which becomes finite as H separates from Hom. 

\subsection{Dynamics in the strongly coupled regime}
Let us now analyze the bifurcation structure of the system in the SC regime. The $(\Delta, S)$-phase diagram plotted in Fig.~\ref{fig3}(a) summarizes the main dynamical regions of the system for $C=5$. One of the main differences is the presence of two distinct, but connected, single-period oscillatory regimes (see red shadowed regions bounded by H$_1$ and H$_2$), which were fused for smaller values of $C$ [see phase diagram in Fig.~\ref{fig2}(a)].
The steady-state bifurcations (i.e., the saddle-nodes) are the same as those already plotted in Fig.~\ref{fig1}(b). Besides the 
two Hopf bifurcations H$_1$ and H$_2$, two SNP$_{1,7}$, several period-doubling bifurcations PD$_{1-3}$, and two homoclinic bifurcations Hom$_{1,2}$ are drawn.

To understand this diagram we take several slices at constant $S$. 
The corresponding bifurcation diagrams are shown in Fig.~\ref{fig3}(i)-(v). Let us first analyze the bifurcation structure and dynamics around the right nonlinear resonance emerging from $C_1$. The modification of the dynamical scenario around this resonance is depicted in Figs.~\ref{fig3}(i)-(iv).

The diagram shown in Fig.~\ref{fig3}(i) [$S=4.5$] intersects H$_2$ at two points, labeled H$_2^{a,b}$, from where periodic oscillations arise. Due to the complexity of this scenario, we have used different colors for each of the limit cycles. In red we plot the maximum and minimum of the oscillation arising from H$_2^a$ and in orange the one emerging from H$_2^b$. The linear stability of these limit cycles is depicted with solid lines for stable states, and dashed lines for the unstable ones. The limit cycle arising from H$_2^a$, hereafter $\Gamma_a$, undergoes a pair of secondary SNP$_{1,2}$. For simplicity, we only plot SNP$_1$ in Fig.~\ref{fig3}(a). Increasing $\Delta$, $\Gamma_a$ encounters the homoclinic bifurcation Hom$_{1}^a$ where it is destroyed.
%
The limit cycle originating from H$_2^b$, however, quickly undergoes a PD$_1$ bifurcation when decreasing $\Delta$.
We label this oscillatory state $\Gamma_b$. Eventually, the extrema of $\Gamma_b$ develop a spiral structure which collapses to the Hom$_{1}^b$ (see close-up view). This spiral behavior is typical of one type of homoclinic bifurcation \cite{Giraldo2020}.
Between Hom$_{1}^a$ and Hom$_{1}^b$ the only attractor of the system is $\psi_e^\alpha$.

With increasing $S$ [see Fig.~\ref{fig3}(a)], the period-doubling bifurcation line PD$_2$ appears, and SN$_2^{l,r}$ 
are created at $C_2$. The bifurcation diagram in this regime is like the one shown in Fig.~\ref{fig3}(ii) for $S=5$. For this value, PD$_2$ is sliced in two points, namely 
PD$_2^{a,b}$
With increasing $\Delta$, $\Gamma_a$ increases its amplitude, and eventually starts to spiral around 
Hom$_{1}^a$, where it is finally destroyed (see close-up view). 
$\Gamma_b$ now emerges from H$_2^b$ subcritically and stabilizes at SNP$_3$ before losing stability in PD$_1$. Once this point is crossed, $\Gamma_b$ describes a large spiral, before dying at Hom$_1^b$. 

\begin{figure}
\includegraphics{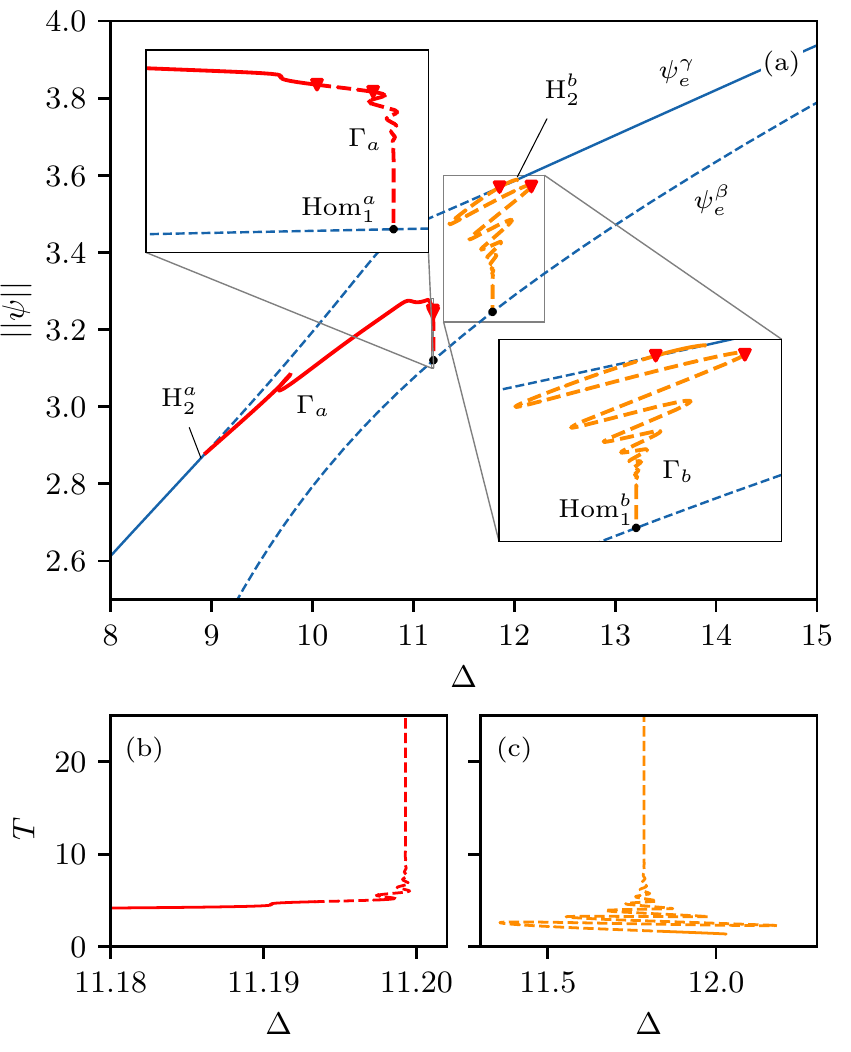}
\caption{Shilnikov homoclinic bifurcations for $C=5$ and $S=4.5$. (a) Close-up view of the diagram shown in Fig.~\ref{fig3}(i) around Hom$_1^{a,b}$. Here we plot $||\psi||$ as a function of $\Delta$. (b) shows the divergence of the period $T$ of $\Gamma_a$ at Hom$_1^a$, and (c) the damped oscillatory modification of the period of $\Gamma_b$ with $\Delta$ around Hom$_1^b$. See Fig.~\ref{fig3} for the definition of the labels.}
\label{fig4}
\end{figure}

Figure~\ref{fig3}(iii) shows the modification of the bifurcation diagram for $S=5.5$.
 As SNP$_{1,2}$ are further apart, we can see that stable oscillations of different amplitudes coexist in a narrow $\Delta$ interval in-between SNP$_{1}$ and the closest PD$_2$ point.
However, the main difference is seen in the right part of the diagram. $\Gamma_b$ emerges subcritically from H$_2^b$ (see orange curve), and becomes stable at SNP$_3$, just before being destabilized at PD$_1$ once more. However, in contrast to Fig.~\ref{fig3}(ii), $\Gamma_b$ does not describe a spiral around Hom$_2^a$ but approaches that point monotonically. Very close to  Hom$_2^a$, the cycle is created again at Hom$_2^b$, leading to the purple unstable curve $\Gamma_c$. This cycle is mainly unstable and spirals around Hom$_1^b$, where it is finally destroyed. A detailed study of this configuration is presented in Sec.~\ref{sec:4} (see Fig.~\ref{fig4}).  

The bifurcation structure is very similar for $S=6$ [see Fig.~\ref{fig3}(iv)], 
with larger
separations between the different saddle-node bifurcations. For simplicity, we have omitted the solution branches associated with $\Gamma_c$. For this value of $S$, Hom$_1^{a,b}$ occurs near one another. From a stability perspective, everything is equivalent to the case for $S=5.5$. 

We have discussed so far the bifurcation structure of the dynamical states emerging from H$_2$ (i.e. around the right resonance) as this bifurcation is first encountered when $S$ is increased. 
However, for $S$ larger than $\approx7.5$, complex nonlinear dynamics also occur around the left nonlinear resonance.
Figure.~\ref{fig3}(v) illustrates a slice of Fig.~\ref{fig3}(a) for $S=8$, where both resonances are plotted. For this value, the bifurcation structure of the oscillations around the tilted right resonance becomes much more complex. Regarding the tilted left one, a new oscillatory state $\Gamma_d$ emerges supercritically from H$_1^a$. $\Gamma_d$ increases its amplitude with $\Delta$ and becomes unstable at SNP$_5$. From H$_1^b$ the periodic orbit also emerges supercritically, although it undergoes SNP$_6$  where it becomes unstable. 
These oscillations correspond to the instability mechanism reported in Refs. \cite{Sarchi2008, Zambon2020} in which the nonlinearity shifts the resonances so as to allow for a resonant four-wave-mixing process with signal and idler photons respectively on the bonding- and antibonding-like mode of the dimer. On the contrary, the parametric instability originating from H$_2$ involves signal and idler photons on both the bonding- and antibonding-like modes (See Supplemental Material of Ref \cite{yelo-sarrion_self-pulsing_2021}).


From each PD bifurcation, a period-doubling cascade can be triggered, potentially leading to chaotic dynamics \citep{ott_chaos_2002}. Period doubled states and chaotic ones will be analyzed in detail in Sec.~\ref{sec:5}.

For $C=5$, neither the Hopf nor the homoclinic  bifurcations emerge from TB points, differently to the case in Sec.~\ref{sec:2} for $C=1.5$. Regarding, H$_{1,2}$ we have found that they extend to large values of $\Delta$ and $S$ (far from the range of applicability of the model), and no signs have been found about their relation with TB points. Similarly, we have established that Hom$_{1}$ forms a close loop in the parameter space, and thus they are detached from any codimension-two point. 


\begin{figure*}[!t]
\includegraphics[scale=0.95]{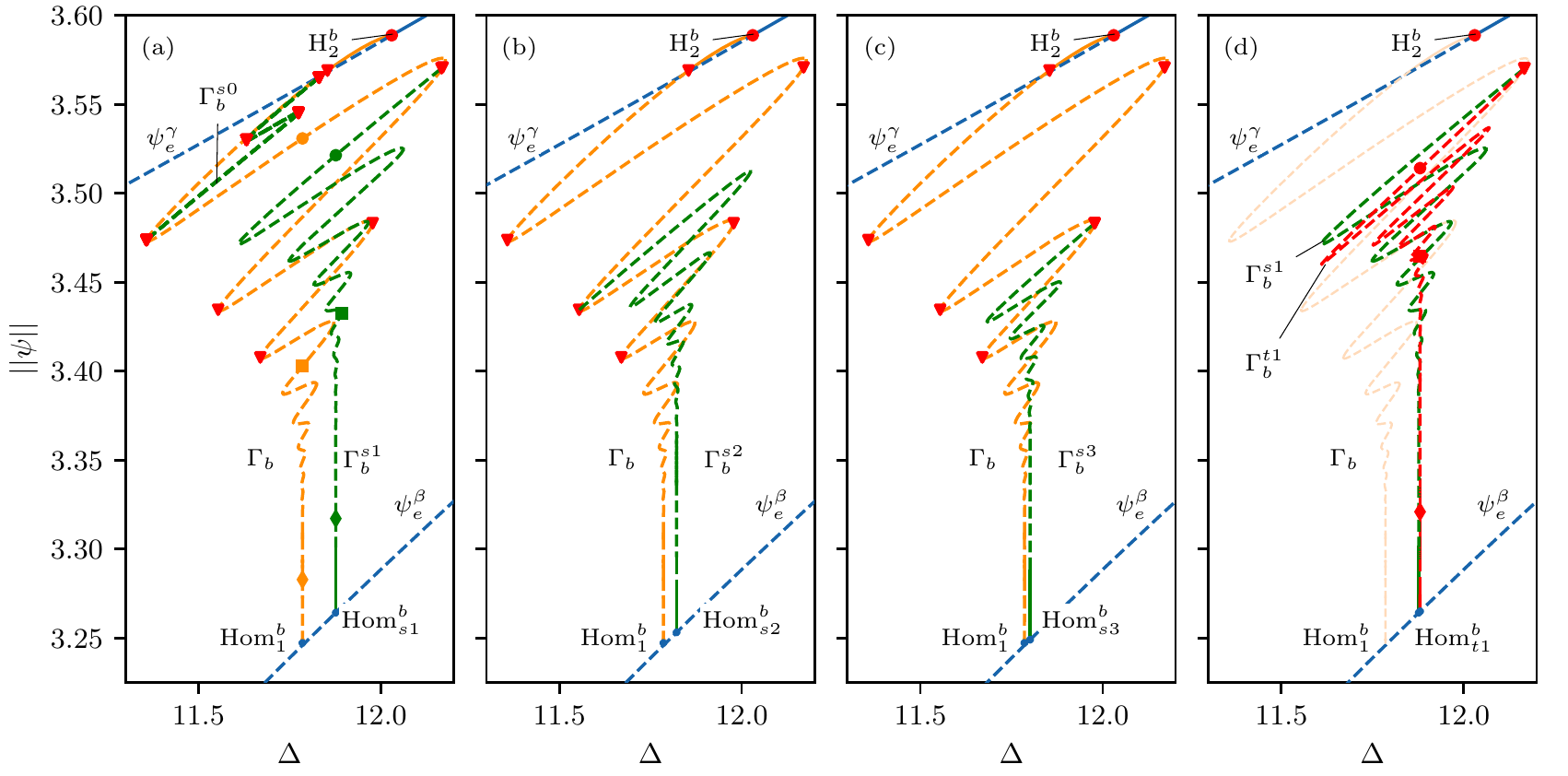}
\includegraphics[scale=0.95]{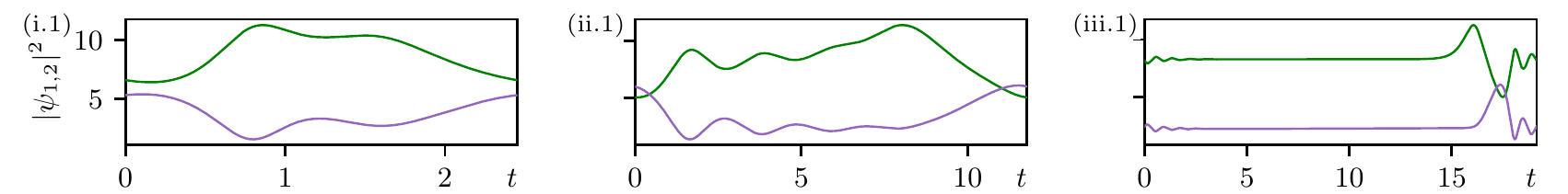}
\includegraphics[scale=0.95]{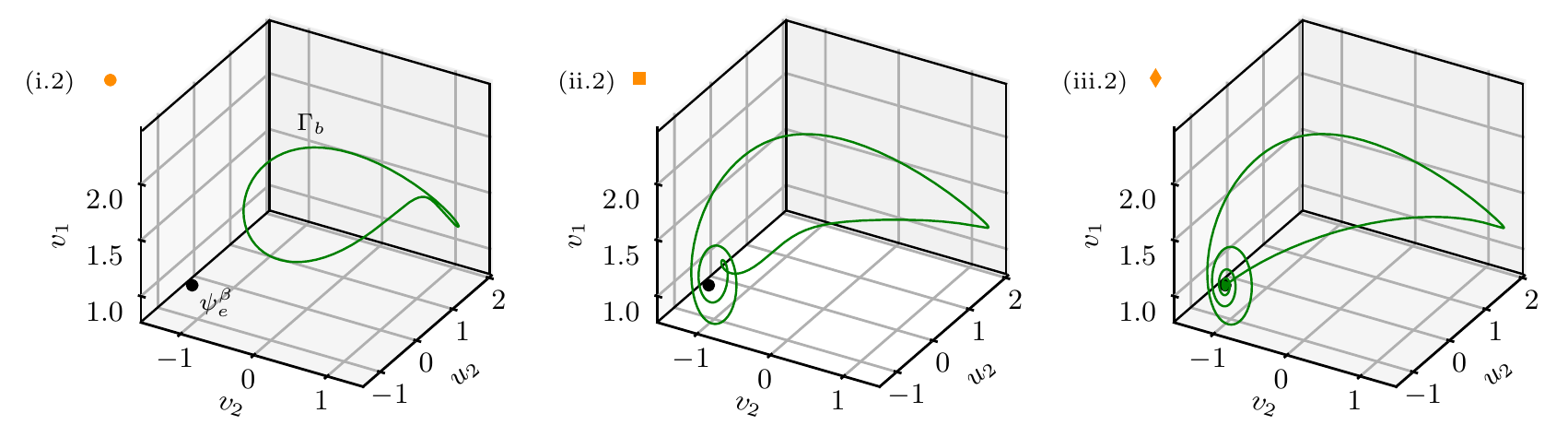}
\caption{Bifurcation diagram of the principal and subsidiary periodic orbits close to Hom$_1^b$ for $C=5$ and $S=4.5$. (a) shows the bifurcation curve associated with the principal orbit $\Gamma_b$ emerging from H$_2^b$ and dying at Hom$_1^b$. Two subsidiary branches $\Gamma_b^{s0}$ and $\Gamma_b^{s1}$, are also plotted. $\Gamma_b^{s1}$ connects with $\psi_e^\beta$ at Hom$_{s1}^{b}$. (b) shows the principal orbit $\Gamma_b$ and the subsidiary curve arising from the fourth PD bifurcation. We label with orbits and its bifurcation curve $\Gamma_b^{s2}$. In (c) we plot $\Gamma_b$ and another subsidiary orbit $\Gamma_b^{s3}$ which dies at Hom$_{s3}^b$. (d) shows the subsidiary curve and $\Gamma_b^{s1}$, and the tertiary oscillatory state $\Gamma_b^{t1}$ arising from it. The modification of the oscillatory state $\Gamma_b$ descending its bifurcation curve [see ${\color{BurntOrange}\bullet},~{\color{BurntOrange}\blacksquare},~{\color{BurntOrange}\blacklozenge}$ in (a)] is depicted in panels (i)-(iii) where the temporal trace and 3D attractors are depicted. }  
\label{fig5}
\end{figure*}

\begin{figure*}[!t]
\includegraphics{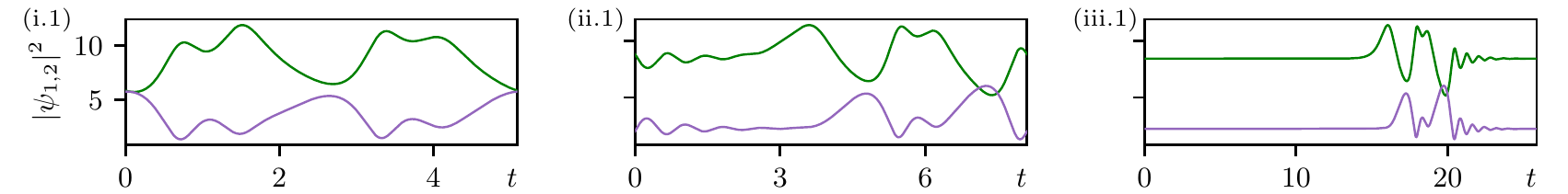}
\includegraphics{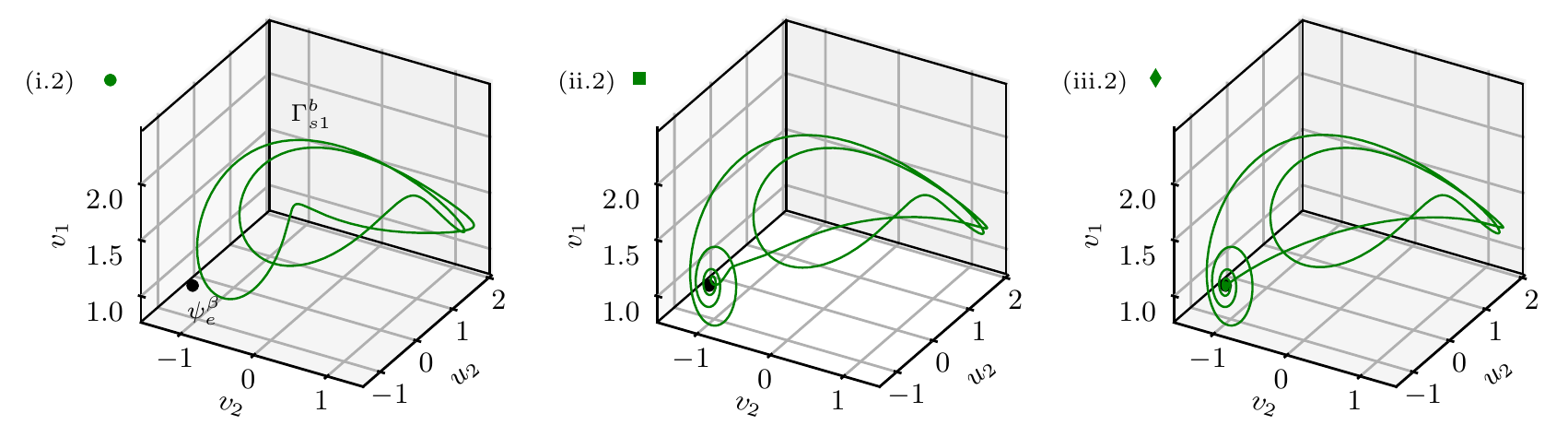}
\includegraphics{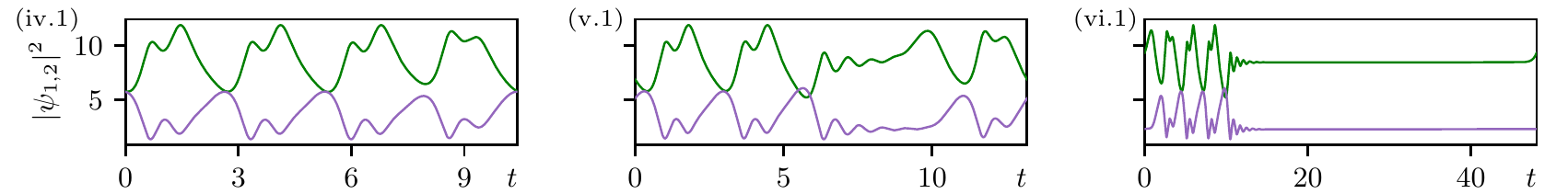}
\includegraphics{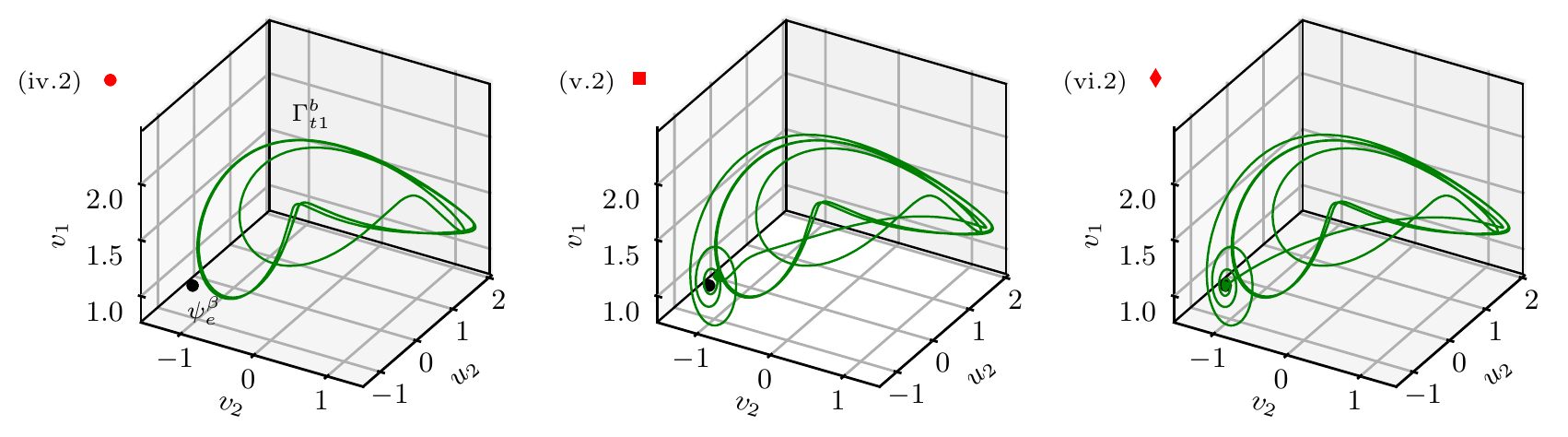}

\caption{Temporal trace and projection on the $\{(v_2,u_2,v_1)\}$-subspace of the subsidiary and tertiary oscillatory states $\Gamma_b^{s1}$ and $\Gamma_b^{t1}$. Panels (i)-(iii) show the modification of $\Gamma_b^{s1}$ along its bifurcation curve in Fig.~\ref{fig5}(a) [see ${\color{OliveGreen}\bullet},~{\color{OliveGreen}\blacksquare},~{\color{OliveGreen}\blacklozenge}$]. At Hom$^b_{s1}$ a 2-homoclinic orbit is created which is very similar to the state shown in (iii). Panels (iv)-(vi) show the modification of $\Gamma_b^{t1}$ [see ${\color{red}\bullet},~{\color{red}\blacksquare},~{\color{red}\blacklozenge}$] along its bifurcation curve while approaching Hom$^b_{t1}$ in Fig.~\ref{fig5}(d). At Hom$^b_{t1}$ a 2-homoclinic orbit is created which is very similar to the state shown in (vi).   }  
\label{fig6}
\end{figure*}

\section{Homoclinic bifurcations}\label{sec:4}
In previous sections, we have found that for some ranges of parameters the periodic oscillations emerging at Hopf bifurcations die out in global homoclinic bifurcations. These bifurcations take place when a limit cycle $\Gamma$ collides with an unstable (hyperbolic) equilibrium for some set of parameters. As the cycle approaches the unstable equilibrium, its period drastically grows, diverging at Hom \citep{glendinning_stability_1994,wiggins_introduction_2003}. At this point, the periodic solution $\Gamma$ becomes a homoclinic orbit $\gamma$, i.e., a closed trajectory linking the unstable equilibrium with itself. Depending on the nature of such equilibrium, different types of Hom bifurcations take place \cite{laing_bifocal_1997,glendinning_stability_1994,glendinning_stability_1994}. In our system we have identified two types corresponding to the following conditions:
\begin{itemize}
    \item When the leading eigenvalues $\lambda_{1,2}$ of the Jacobian $\mathcal{J}$ are real (i.e. $\lambda_{s,u}=a_{s,u}\in\mathbb{R}$) such that $a_s<0<a_u$, the Hom orbit $\gamma$ is biasymptotic to a saddle equilibirum, and the Hom bifurcation is commonly called {\it saddle-loop} Hom bifurcation \citep{izhikevich_neural_2012}. 
   
\item If the leading eigenvalues of $\mathcal{J}$ are one real and one complex conjugate pair (i.e., $\lambda_s=a_s\pm i\omega_s$ and $\lambda_u=a_u$), with $a_s<0<a_u$, and $\omega_s>0$, the Hom orbit is biasymtotic to a saddle-focus equilibrium, and the bifurcation is known as saddle-focus Hom bifurcation or {\it Shilnikov} bifurcation \cite{homburg_homoclinic_2010}. 
A relevant parameter describing the nature of these points is the {\it saddle-index} quantity \cite{glendinning_stability_1994} 
\begin{equation}
\delta\equiv -{\rm Re}[\lambda_s]/\lambda_u.
\end{equation}
When $\delta>1$, the saddle-focus homoclinic orbit is said to be {\it tame} \citep{homburg_homoclinic_2010} and the dynamics are essentially the same as in the saddle-loop case. In contrast, when $\delta<1$ the homoclinic orbit is called {\it wild} and the dynamics of the system around it is richer. In particular, there is an infinite number of SNP and period-doubling bifurcations in any parameter interval containing the bifurcation \citep{homburg_homoclinic_2010,glendinning_stability_1994,glendinning_local_1984}. 


\end{itemize}
Let us analyze these bifurcations in our case.


\subsection{Saddle-focus (Shilnikov) homoclinic bifurcation}
Most of the homoclinic bifurcations and orbits appearing in our system are of the Shilnikov type. The bifurcation diagram plotted in  Fig.~\ref{fig4}(a) shows the appearance of these bifurcations for $C=5$ and $S=4.5$. It consists of a close-up view of the diagram shown in Figs.~\ref{fig3}(i). For more clarity, we plot the $L_2$-norm
\begin{equation}
||\psi||\equiv \sqrt{T^{-1}\int_0^T(|\psi_1(t)|^2+|\psi_2(t)|^2)dt,}
\end{equation}
 as a function of $\Delta$, where $T$ is the period of the oscillatory state. This allows us to better visualize the occurrence of the different Hom bifurcations. 
 
 Let us first take a look at the self-pulsing state $\Gamma_b$ emerging from H$_2^b$. Soon after its birth,  $\Gamma_b$ undergoes a first PD bifurcation, and its norm  $||\psi||$ changes in a damped oscillatory fashion while approaching asymptotically Hom$_{1}^b$ [see Fig.~\ref{fig4}(a)]. This structure corresponds to the spiral shown in the inset of Fig.~\ref{fig3}(i), and  each fold to an SNP.  All along this curve, the period of $\Gamma_b$ increases as approaching Hom$_{1}^b$, and in doing so, it describes the damped oscillatory curve in $\Delta$ plotted in Fig.~\ref{fig4}(c). We refer to this state as {\it primary periodic} orbit \cite{glendinning_local_1984}.
 
 
 In Fig.~\ref{fig5}(a) we plot the bifurcation curve associated with the principal orbit $\Gamma_b$. Its modification along such a curve is depicted in Fig.~\ref{fig5}(i)-(iii). In Fig.~\ref{fig5}(i.1) we show the temporal trace of $\Gamma_b$ during one oscillatory period, and in Fig.~\ref{fig5}(i.2) its 3D representation in the phase subspace $\{(v_2,u_2,v_1)\}$. 
 
 Moving down along this diagram, the period $T$ of $\Gamma_b$ increases [see Fig.~\ref{fig5}(ii.1)], while the periodic attractor approaches the saddle-focus equilibrium $\psi_e^\beta$. In doing so, the periodic orbit temporarily follows the flow around $\psi_e^\beta$, leading to the almost spiral-like trajectory shown in Fig.~\ref{fig5}(ii.2). Close to Hom$_1^b$, the periodic orbit looks like the one shown in Figs.~\ref{fig5}(iii), where the oscillatory period has considerably increased, and where the orbit describes a spiral trajectory around $\psi_e^\beta$. The behavior of the trajectory around this point follows the unstable and stable manifolds of the SF equilibrium, being the latter one, responsible of the oscillatory tail shown in its temporal trace. 
 
 Approaching Hom$_{1}^b$, the period of $\Gamma_b$ tends to infinite, and at that point, $\Gamma_b$ collides with the SF equilibrium $\psi_e^\beta$, leading to the formation of the wild Shilnikov homoclinic orbit $\gamma_b$. This homoclinic orbit is very similar to the periodic orbit shown in Figs.~\ref{fig5}(iii), however, for this set of parameters, it is unstable. The eigenvalues and saddle-index associated with this point are shown in Table~\ref{tabla1}.

\begin{table}[!t]
\begin{tabular}{ | c | c | c | c | c | c |}
  \hline
    \hline
     Label  & $\psi_e$-type  & $\Delta$  & $\lambda_u$ & $\lambda_s$ &$\delta$ \\
  \hline
  Hom$_1^b$& SF & 11.7857 & 2.4078 & -1 - $i$7.9829j & 0.4153 \\
  
  Hom$_{s1}^{b}$& SF & 11.8769 & 2.4282 & -1 - $i$8.0693 & 0.4118 \\
 
  Hom$_{s2}^{b}$ & SF & 11.8216 & 2.4159 & -1 - $i$8.0166 & 0.4139 \\

  Hom$_{s3}^{b}$ & SF & 11.8009 & 2.4113 & -1 - $i$7.9971& 0.4147 \\
  
  Hom$_{t1}^{b}$ & SF & 11.8804 & 2.4289 & -1 - $i$8.0727& 0.4116 \\

  \hline
  \hline
\end{tabular}
\caption{Features and relevant information about the wild Shilnikov homoclinic bifurcations plotted in Fig.~\ref{fig5}. SF stands for saddle-focus equilibrium, $\lambda_s$ represents the stable eigenvalues, $\lambda_u$ is unstable real eigenvalue, and $\delta$ is the saddle-index associated with the homoclinic bifurcation.}
\label{tabla1}
\end{table}
 Very close to the SNPs, the primary bifurcation curve undergoes PD bifurcations [ see {\color{red}$\blacktriangledown$} in Fig.~\ref{fig5}(a)], from where other {\it secondary} or {\it subsidiary} orbits emerge. The bifurcation curves associated with two of those secondary orbits $\Gamma_b^{s0}$ and $\Gamma_b^{s1}$ are plotted in green in Fig.~\ref{fig5}(a).
 
The modification of the period-2 orbit $\Gamma^{s1}_b$ around this diagram is shown in Fig.~\ref{fig6}(i)-(iii). This orbit is well illustrated in Fig.~\ref{fig6}(i.2). As in the single-period case, the period diverges as we descend the green diagram [see Fig.~\ref{fig6}(ii.1) ] and approach $\psi_e^\beta$. The vicinity of this equilibrium leads to the characteristic spiral trajectory shown in Fig.~\ref{fig6}(ii.2). Further decreasing $||\psi||$, $\Gamma_b^{s1}$ approaches Hom$_{s1}^{b}$, where it is destroyed and the homoclinic orbit $\gamma_b^{s1}$ is created. This state is known as a {\it 2-homoclinic orbit} and is very similar to the orbit plotted in Fig.~\ref{fig6}(iii).
 
 
 Similarly, secondary orbits of larger period arise from each of the PD bifurcations as one proceeds down in the diagram. Two of these curves are plotted in Figs.~\ref{fig5}(b) and (c).  Decreasing $||\psi||$, these orbits die at Hom$_{s1}^{b}$ and Hom$_{s2}^{b}$, leading to new Shilnikov homoclinic orbits which occurs very close to Hom$_{1}^{b}$. The characteristics of these orbits are also shown in Table~\ref{tabla1}. 
 This phenomenon is known as {\it homoclinic doubling cascade} \cite{homburg_homoclinic_2010}, and has been analyzed numerically in \cite{oldeman_death_2000}. The main idea is that
for $\Delta>\Delta_{\mathrm{Hom}_1^{b}}$, an infinite number of $N$-homoclinic orbits $\gamma_N$ (with $N>0$) accumulate on the right side of the primary branch finishing at $\mathrm{Hom}_1^{b}$.
 
The secondary green bifurcation curves $\Gamma_b^{s1,2,3}$ does also undergo PD bifurcation from where period-4 orbits emerge, leading to similiar homoclinic doubling cascades. We represent this orbit as $\Gamma_b^{t}$, where the subindex $t$ stands for {\it terciary}. The bifurcation curve associated with $\Gamma_b^{t1}$ is shown in  Fig.~\ref{fig5}(d) together with $\Gamma_b^{s1}$. The modification of $\Gamma_b^{t1}$ along this diagram is shown in Figs.~\ref{fig6}(iv)-(vi). 
 As proceeding down in the diagram, the modification of the orbits is similar to the single and period-2 cases. At Hom$_{t1}^{b}$ a 4-homoclinic orbit is formed. This orbit is similar to the long period limit cycle plotted in Fig.~\ref{fig6}(vi).

The limit cycle $\Gamma_a$ emerging from H$_2^a$ [see left red curve in Fig.~\ref{fig4}(a)], also undergoes an oscillatory damped structure around Hom$_{1}^a$. However, the oscillations in $\Delta$ are more damped. The period of $\Gamma_a$ diverges as approaching Hom$_{1}^a$, following the same oscillatory tendency [see Fig.~\ref{fig4}(b)]. At Hom$_{1}^a$, the period of the oscillations becomes infinite, and $\Gamma_a$ becomes the homoclinic orbit $\gamma_a$.

\subsection{Saddle-Loop homoclinic bifurcation}
\begin{figure}
\includegraphics{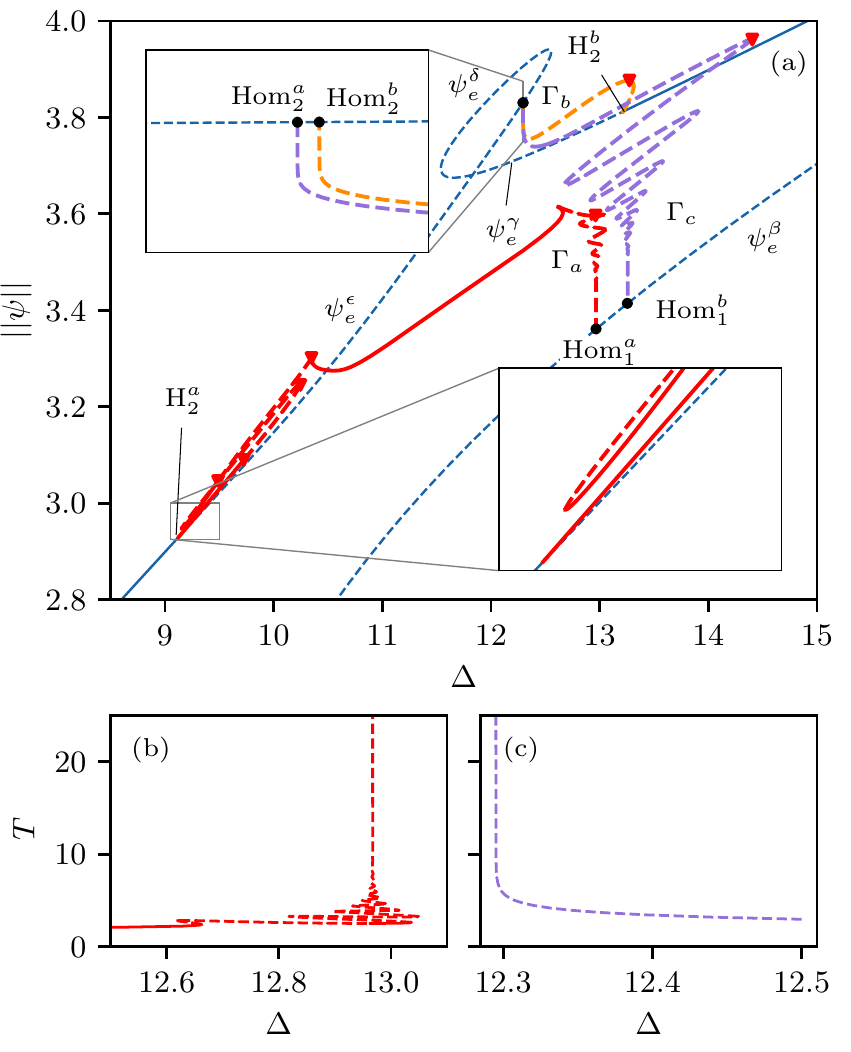}
\caption{(a) Bifurcation diagram showing $||\psi||^2$ as a function of $\Delta$ around the Shilnikov bifurcations Hom$_1^{a,b}$ and saddle-loop bifurcations Hom$_2^{a,b}$ for $C=5$ and $S=5.5$. (b) shows the divergence of the period of $\Gamma_a$ around Hom$_1^a$. (c) shows the monotonic divergence of the period of $\Gamma_c$ when approaching Hom$_1^a$. }
\label{fig8}
\end{figure}

Increasing $S$, the previous scenario is modified as illustrated in Fig.~\ref{fig8}(a) for $S=5.5$.
Here, Hom$_{1}^{a,b}$ are closer to one another. Furthermore, close to H$_2^a$, $\Gamma_a$ undergoes a series of SNP and PDs, which were absent before. The divergence of the period close to  Hom$_{1}^{a}$ is depicted in Fig.~\ref{fig8}(b).

\begin{table}[!t]
\begin{tabular}{ | c | c | c | c | c | c |}
  \hline
    \hline
     Label  & $\psi_e$-type  & $\Delta$  & $\lambda_u$ & $\lambda_s$ &$\delta$ \\
  \hline
  Hom$_2^a$& S & 12.2951  & 1.1762 & -3.1762 & 2.7003 \\
  
  Hom$_2^b$ & S & 12.2953 & 1.1134 & -3.1134 & 2.7962 \\
 
  Hom$_1^a$ & SF & 12.9672 & 3.4932 & -1  - $i$9.7634& 0.28626 \\

  Hom$_1^b$ & SF & 13.2552 & 3.5641 & -1 - $i$1.0048& 0.28057 \\
  \hline
  \hline
\end{tabular}
\caption{ Features and relevant information about the homoclinic bifurcations are plotted in Fig.~\ref{fig8}. SF stands for saddle-focus equilibrium, S corresponds to a saddle equilibrium, $\lambda_s$ represents the stable eigenvalues, $\lambda_u$ is an unstable real eigenvalue, and $\delta$ is the saddle-index associated with the homoclinic bifurcation.
}
\label{tabla2}
\end{table}

In contrast to the situation shown for $S=4.5$, the bifurcation curve $\Gamma_c$ emerging from Hom$_{1}^{b}$ (in purple) does not connect with H$_2^b$, but with a new homoclinic bifurcation Hom$_2^a$ taking place at the saddle equilibrium $\psi_e^\epsilon$. In this case [see a close-up view in Fig.~\ref{fig8}(a)] the reconnection follows a monotonic growth in $||\psi||$ very different from the oscillatory one shown in the saddle-focus case. This saddle-loop homoclinic bifurcation, labeled Hom$_{2}^a$, is characterized by a scaling law  $T\propto-{\rm ln}(\Delta-\Delta_{{\rm Hom}_2^a})/a_u$, which governs the period of the oscillatory state very close to the bifurcation \citep{glendinning_stability_1994}. This divergence is plotted in 
Fig.~\ref{fig8}(c).


The limit cycle $\Gamma_b$ emerging from H$_2^b$ does not die at Hom$_{1}^b$, but at another saddle-loop homoclinic bifurcation Hom$_2^b$. These homoclinic orbits are unstable, and therefore cannot be observed in direct numerical time simulations.

\begin{figure*}[!t]
\includegraphics[scale=0.87]{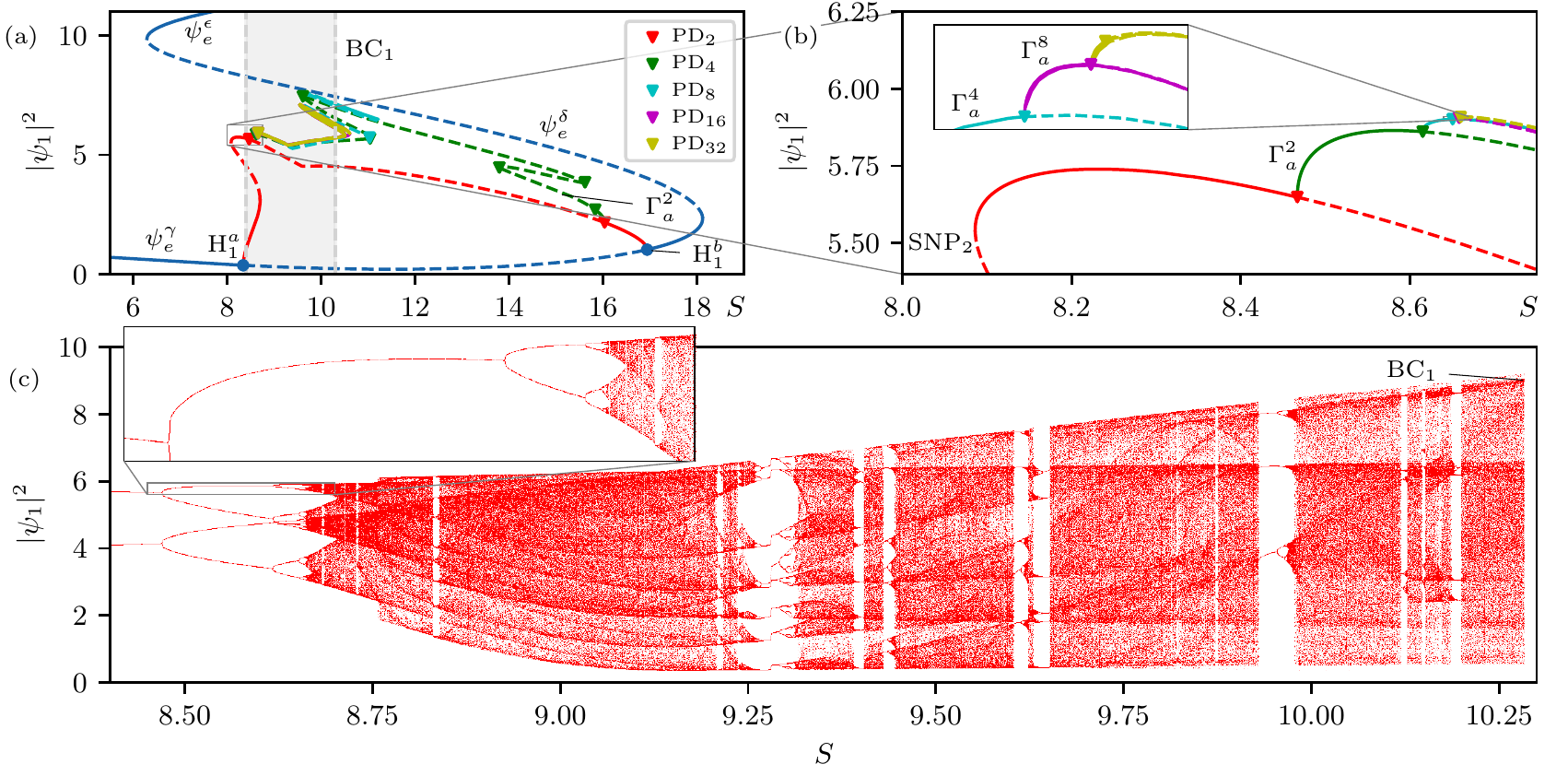}
\includegraphics[scale=0.87]{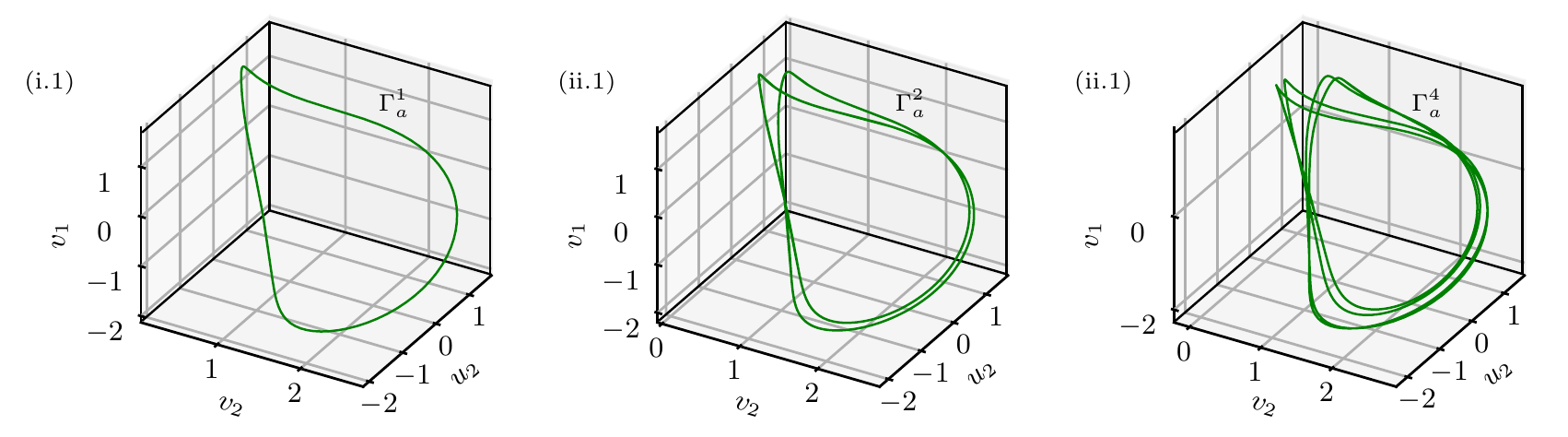}
\includegraphics[scale=0.87]{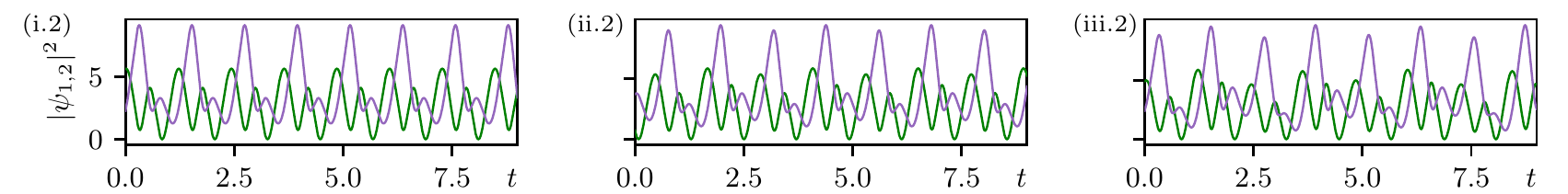}
\includegraphics[scale=0.87]{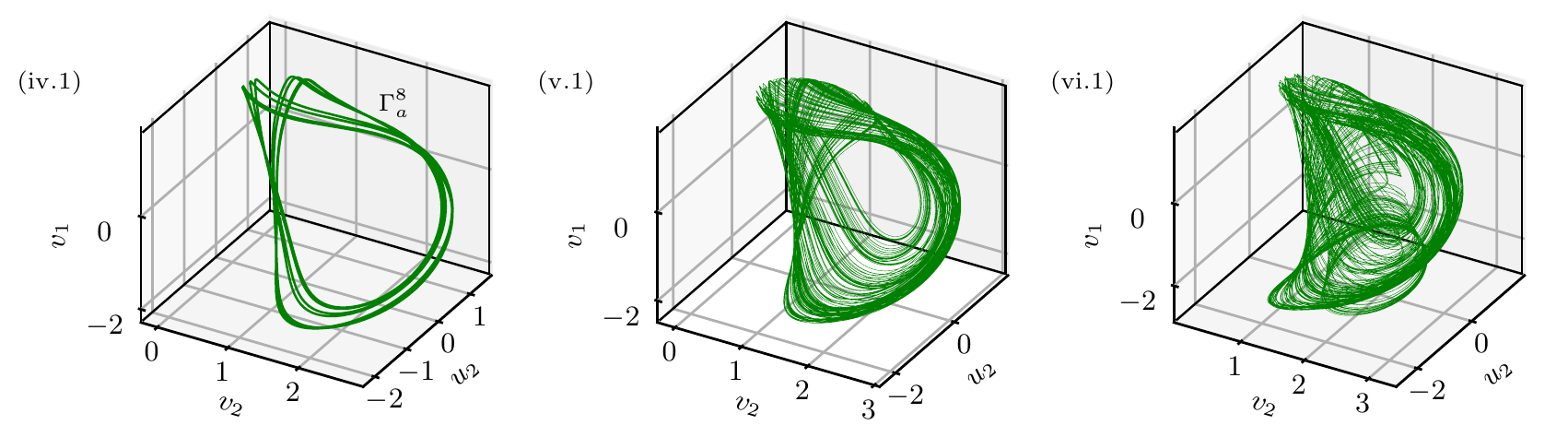}
\includegraphics[scale=0.87]{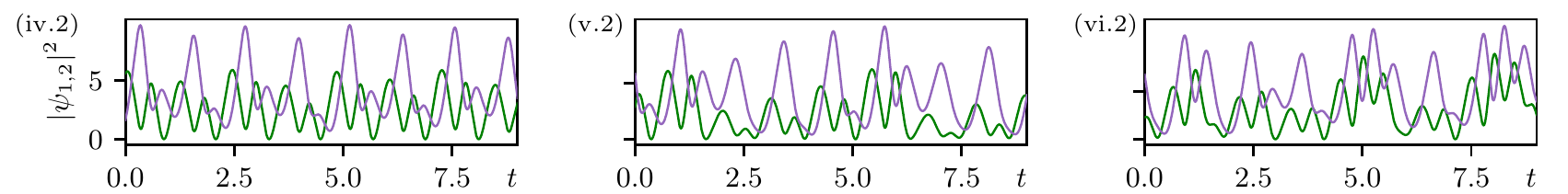}
\caption{Period doubling cascade and route to chaos for $C=\Delta=5$. (a) shows the modification of the norm $|\psi_1|^2$ as a function of $S$. The different colored branches correspond to the oscillatory states with different periodicities. We mark the Hopf bifurcations H$_1^{a,b}$ and different PD bifurcations. Solid (dashed) lines correspond to stable (unstable) states. Panel (b) is a close-up view of (a) around the period-doubling cascade. In (c) we plot the Feigenbaum diagram associated with the shadowed gray in (a), which shows the local maxima and minima modification of the dynamical attractors with changing $S$. Panels (i)-(vi) show the time trace (top panel) and projection of the attractors on the $\{(v_2,u_2,v_1)\}$-subspace (bottom panels) for different values of $S$.  
From (i) to (vi) these values are respectively $S = 8.45, 8.50, 8.64, 8.84, 9.10, 10.26$. (v) and (vi) are chaotic attractors. }  
\label{fig9}
\end{figure*}

\section{Chaotic dynamics}\label{sec:5}
In this section, we analyze the emergence of chaotic dynamics, one close to $\Delta=C$, the other when $\Delta\gg C$, finding two main scenarios leading to chaos. The first one involves a period-doubling cascade \citep{ott_chaos_2002}, while the second one is associated with the presence of a homoclinic bifurcation \citep{wiggins_introduction_2003}.

 \begin{figure*}
\includegraphics{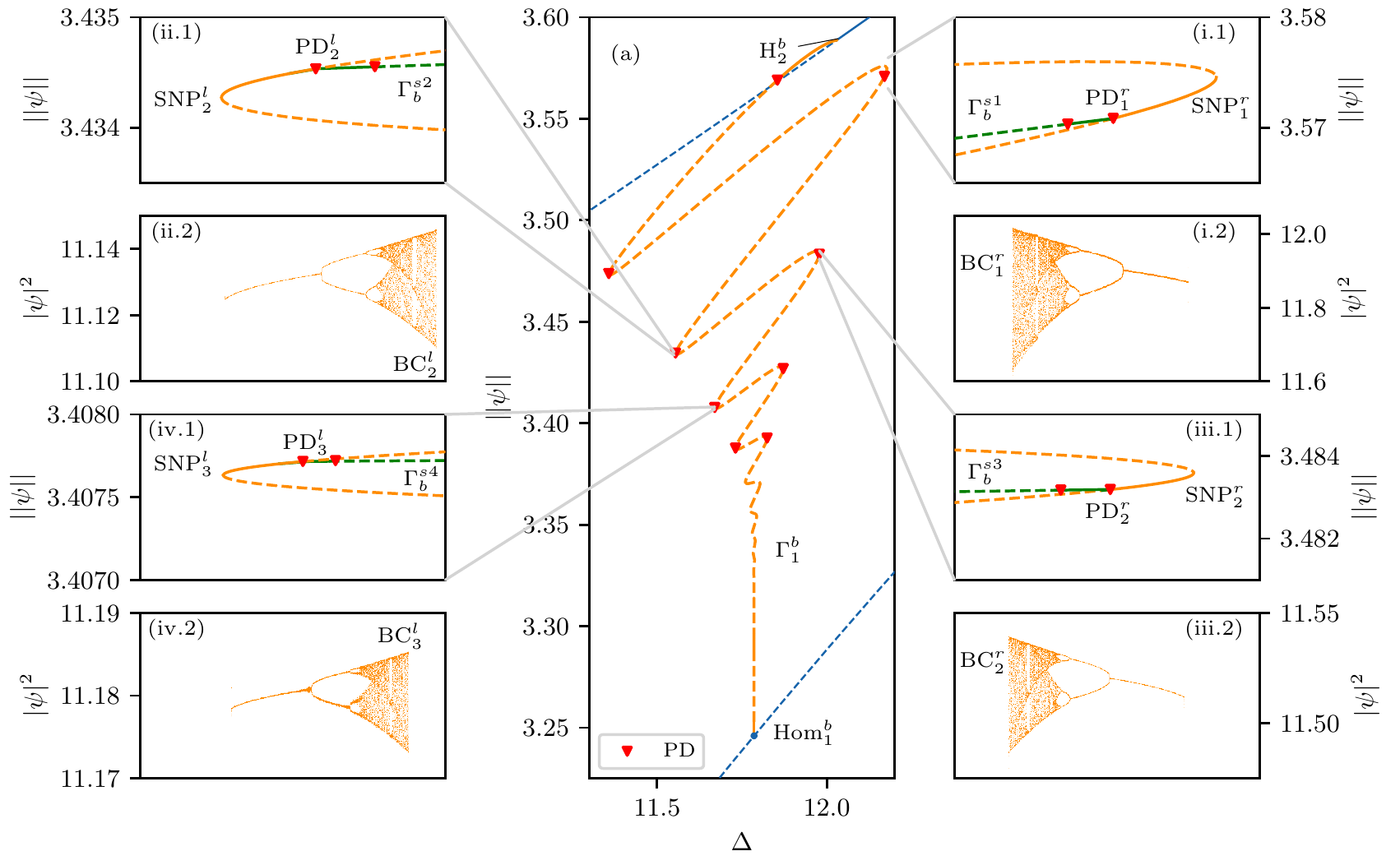}
\caption{Period doubling route to chaos near a wild Shilnikov homoclinic orbit. (a) shows the bifurcation curve of the principal periodic orbit $\Gamma_b$ emerging from H$_2^b$ and dying at Hom$_1^b$. Panels (i.1)-(iv.1) show a close-up view of (a) around  SNP$_1^{r}$, SNP$_2^{l}$, SNP$_2^{r}$ and SNP$_3^{l}$, and the subsidiary dynamical state branches $\Gamma_b^{s1-s4}$. Panels (i.2)-(iv.2) show the Feigenbaum diagram and standard period-doubling route to chaos corresponding to the same branches. The chaotic states die at several boundary crises BC$_j^{l,r}$. }  
\label{fig10}
\end{figure*}

\subsection{Period doubling cascade to chaos}\label{sec:5a}
As shown previously, the periodic orbits emerging from the different H bifurcations may encounter PD bifurcations, where  period-2 orbits are created, while the former ones become unstable. This PD process might repeats in cascade and then lead to temporal chaos. 
Figure~\ref{fig9}(a) shows the occurrence of these bifurcations in a diagram which depicts $|\psi_1|^2$ as a function of $S$ for $C=\Delta=5$. This diagram corresponds to a vertical slice of the phase diagram shown in Fig.~\ref{fig3}(a). We only plot the three equilibrium branches $\psi_e^\gamma$, $\psi_e^\delta$, and $\psi_e^\epsilon$, which are linked through the bifurcations SN$_3^l$ and SN$_3^r$. The Hopf line H$_1$ cuts the stable branch at two points that we label H$_1^{a,b}$. From these points, the single period limit cycle $\Gamma_a$ appears supercritically. The periodic attractor associated with this orbit is plotted in the 3D phase subspace $\{(v_2,u_2,v_1)\}$ depicted in Fig.~\ref{fig9}(i.1). The temporal trace is shown in Fig.~\ref{fig9}(i.2).


On the left, this state undergoes a pair of SNPs where it loses and gains stability. Once SNP$_2$ is passed, $\Gamma_a$ remains stable until PD$_2$. This situation is shown in the close-up view plotted in Fig.~\ref{fig9}(b). At PD$_2$, an oscillatory state with two different periods, hereafter $\Gamma_a^2$, emerges and remains stable until PD$_4$. An example of this state is plotted in Fig.~\ref{fig9}(ii). From PD$_4$ a new oscillatory state, $\Gamma_a^4$, emerges [see Figs.~\ref{fig9}(iii)] and after that, a cascade of period-doubling bifurcations (PD$_8$, PD$_{16}$, etc) occurs in a very short interval of $S$. From these bifurcations, the states $\Gamma_a^8$ shown in Figs.~\ref{fig9}(iv), and $\Gamma_a^{16}$ (not shown here), emerge. Increasing a bit further $S$, the system reaches a regime characterized by chaotic states like the one shown in Fig.~\ref{fig9}(v).

After passing PD$_2$, $\Gamma_a^2$ undergoes several SNPs and PDs for increasing values of $S$, and eventually, it connects back to $\Gamma_a$. The different limit cycles undergo several SNPs as shown in the close-up view in Fig.~\ref{fig9}(a). 
 A similar structure is found on the right part of the diagram close to H$_1^b$.


 The period-doubling cascades are also illustrated through the {\it Feigenbaum diagram} \cite{ott_chaos_2002} plotted in Fig.~\ref{fig9}(c). This diagram has been computed by scanning the stable attractors of the system as a function of $S$ and collecting the local maxima and minima of the oscillatory states. The extension of this diagram corresponds to the shadowed gray box in Fig.~\ref{fig9}(a). The close-up view corresponds to the range plotted in 
 Fig.~\ref{fig9}(b).
 
 Increasing $S$, the chaotic attractor increases its morphological complexity and undergoes the typical windows of odd period oscillations \cite{ott_chaos_2002}. 
 A complete understanding of these modifications, and the {\it crisis} suffered by the attractor, requires the analysis of its return map, as reported in \citep{rossler_chaos_1977,barrio_qualitative_2009}. The chaotic nature of this system can be also characterized through the computation of the Lyapunov exponent and the Kaplan-Yorke dimension associated with the dynamics. This type of approach has been applied to investigate the route to chaos in a plasmonic dimer \cite{ziani_investigating_2020}. 
 Such analyses are beyond the scope of this work.
 
The chaotic dynamics persist until a critical value of $S$, where the chaotic attractor collides with an unstable periodic orbit on its basing boundary and is destroyed. This is a typical phenomenon in chaotic dynamics and it is known as {\it boundary crisis} \cite{ott_chaos_2002}. We label this boundary crisis BC$_1$ as depicted in Fig.~\ref{fig9}(a) and (c). After crossing this point,  the only attractor of the system is the stable steady-state $\psi_e^\epsilon$. Similarly, this route to chaos arises from H$_1^b$ on the right and ends in a second BC$_2$ (not shown).

\subsection{Chaos close to a Shilnikov homoclinic bifurcation}\label{sec:5b}
In this section, we analyze the emergence of chaotic dynamics close to homoclinic orbits. 
As stated by the Shilnikov theorem, if the Shilnikov bifurcation is wild (i.e. $\delta<1$), chaotic dynamics is expected in the neighborhood of the homoclinic orbit \cite{homburg_homoclinic_2010}.
The interplay between chaotic dynamics and homoclinic orbits has been studied by different authors, in particular in the context of the Rossler model \cite{arneodo_oscillators_1982,gaspard_what_1983,glendinning_local_1984,malykh_homoclinic_2020}. 

To illustrate this phenomenon here, let us take a look at the curve shown in Fig.~\ref{fig10}, which corresponds to the primary orbit $\Gamma_b$ emerging from H$_2^b$ and dying at Hom$_1^b$. This diagram is a detailed version of the one plotted in Figs.~\ref{fig4}(a) and \ref{fig5}(a). The close-up views of the diagram around the first four SNPs are shown in the insets, together with the PD bifurcations PD$_j^{l,r}$, and the period-2 secondary branches emerging from them. These plots show a similar structure for the left and right folds and extend along with the whole diagram with decreasing $||\psi||$ (not shown here for simplicity). The panels below show the Feigenbaum diagrams for the same interval in $S$ and the period-doubling cascade. The agreement between both diagrams on the onset for the period-2 bifurcations is excellent.  
With increasing $S$, each of these attractors undergoes a BC, and they disappear. 

All along this diagram, SNP$_{i}^{l,r}$ and BC$_{i}^{l,r}$ accumulate asymptotically  around Hom$_1^b$, and so do the chaotic regions. Eventually, the left and right BCs may collide, leading to the merging of the left and right chaotic attractors very close to Hom$_1^b$. Furthermore,  while decreasing $||\psi||$ and approaching Hom$_1^b$, the chaotic attractor comes closer and closer to the SF $\psi_e^\beta$, and the trajectories must start to spiral around this point following the stable and unstable manifolds of $\psi_e^\beta$, in a similar manner as the one shown in Fig.~\ref{fig5} and Fig.~\ref{fig6} for the homoclinic orbits. The type of chaos associated with these attractors is commonly known as {\it spiral chaos} \cite{gaspard_what_1983,malykh_homoclinic_2020}. However, due to the exponentially shrinking of the chaotic intervals when approaching Hom$_1^b$ we have not been able to confirm these two hypotheses. 


\section{Discussions and Conclusions}\label{sec:6}

In this paper, we have presented a systematic study of the temporal dynamics arising in the asymmetrically-driven dissipative photonic Bose-Hubbard dimer model. This model has proved to describe excellently the self-pulsing dynamics of two coupled 
cavities \cite{yelo-sarrion_self-pulsing_2021}.
The particularity of this system is that only one cavity is driven. This asymmetry in the driving leads to an absence of equivalent states (emerging from Pitchfork bifurcations) which are present in symmetrically driven cavities. \citep{Giraldo2020}

Applying methods of dynamical systems and bifurcation theory we have presented a detailed collection of results describing the temporal dynamics of this system in different regimes of operation. 

After introducing the model in Sec.~\ref{sec:1}, we have analyzed the modification of the resonances of the cavity, i.e. the homogenous states of the system, as a function of the driving field amplitude $S$, for two different coupling regimes (see Sec.~\ref{sec:2}). We have referred to these regimes as weakly and strongly coupled regimes (WC and SC, respectively). The next step in this study has consisted in analyzing the linear stability of the homogeneous equilibrium points $\psi_e$ against small perturbations. This analysis has been performed numerically using AUTO-07p \cite{Doedel2009}. The main steady-state bifurcations are depicted in Fig.~\ref{fig1}. Through this analysis, we have detected Hopf bifurcations where $\psi_e$ states are destabilized in favor of periodically oscillating ones. The main dynamical regimes of the system for the WC regime have been summarized in Fig.~\ref{fig2}.

The continuation of oscillatory states  has also been performed using AUTO-07p, as well as the computation of their stability \cite{Doedel2009}. This analysis has led to identifying PD cascades leading to limit cycles with different even periods. The periodic states may also undergo homoclinic bifurcations where they are destroyed. 

We have proceeded similarly for the SC coupling regime ($C=5$), and our findings show a much more complex dynamical scenario which is depicted in Fig.~\ref{fig3}. In contrast to the WC case, here we have found two distinct oscillatory regimes, each one appearing close to one of the two nonlinear resonances of the cavity and [see Fig.~\ref{fig3}(v)] For this value of $C$, these two regimes are disconnected. We have also shown that the limit cycles emerging from the right resonance encounter a variety of homoclinic bifurcations where they die. 

In Sec.~\ref{sec:4} we performed a systematic study of the homoclinic bifurcations appearing in the SC regime. To do so we have applied well-known results of dynamical systems theory \citep{glendinning_stability_1994}. We have identified two main types of homoclinic. One of them is a closed orbit bi-asymptotic to a saddle-focus equilibrium (i.e., Shilnikov homoclinic orbits), while the other one connects a saddle point with itself. The main dynamical implications of both scenarios are described in detail.  

Finally, we have also analyzed the chaotic behavior emerging in the system (see Sec.~\ref{sec:5}). Chaotic dynamics emerge through several period-doubling cascades which occur either close to a homoclinic bifurcation or far from it. The period-doubling cascade and its dynamics has been analyzed using continuation algorithms and direct numerical simulations. One example of the bifurcation structure of such states is presented in Fig.~\ref{fig9} (see Sec.~\ref{sec:5a}) together with a 3D representation of their different attractors.

We have also analyzed the emergence of chaotic dynamics close to the Shilnikov homoclinic bifurcations (see Sec.~\ref{sec:5b}). The main results are depicted in Fig.~\ref{fig10}.
In any case, the period-doubling cascade follows the typical Feigenbaum diagram \citep{ott_chaos_2002}.

The complexity of the dynamics appearing in this model, and the fidelity of our model to describe coupled Kerr cavities suggest that those dynamical regimes may be reachable experimentally. Hence, we hope that these results will be relevant for experimentalists working on these types of systems. 

\section*{Acknowledgements}
This work was supported by the Fonds de la Recherche Scientifique - FNRS under grant No PDR.T.0104.19 and the European Research Council (ERC) under the European Union’s Horizon 2020 research and innovation program (grant agreement No 757800). F.L. and P.P.-R. acknowledge the support of the Fonds de la Recherche Scientifique-FNRS). P. P. -R acknowledges support from the European Union’s Horizon 2020 research and innovation programme
under the Marie Sklodowska-Curie grant agreement no. 101023717.

\bibliography{biblio}
\end{document}